**Title: Potential gains of long-distance trade in electricity**

**Authors: Javier López Prol [a] [*], Karl W. Steininger[b], Keith Williges[b], Wolf D. Grossmann[b], Iris Grossmann[c,d]**

**Declarations of competing interests:** none

[a] Yonsei University Mirae, Department of Economics and Department of Environmental Finance. 1 Yeonsedae-gil, Heungeop-myeon, Wonju-si, Gangwon-do, South Korea.
[*] Corresponding author: email: jlprol@yonsei.ac.kr.
[b] Wegener Center for Climate and Global Change, University of Graz, Brandhofgasse 5, Graz, Austria.
[c] Falk School of Sustainability, Chatham University, Woodland Road, Pittsburgh, United States.
[d] Carnegie Mellon University, Pittsburgh, United States.

**Title: Potential gains of long-distance trade in electricity**

**Abstract:**

Electrification of all economic sectors and solar photovoltaics (PV) becoming the lowest-cost electricity generation technology in ever more regions give rise to new potential gains of trade. We develop a stylized analytical model to minimize unit energy cost in autarky, open it to different trade configurations, and evaluate it empirically. We identify large potential gains from interhemispheric and global electricity trade by combining complementary seasonal and diurnal cycles. The corresponding high willingness to pay for large-scale transmission suggests far-reaching political economy and regulatory implications.

## 1. Introduction

Economic development closely relies on the energy basis it draws upon. The success of modern economic growth arose from the ability to tap into new forms of energy. James Watt´s steam engine allowed for harnessing the potential of coal, the internal combustion engine to run on petroleum, and finally the gas turbine to draw upon natural gas (Sachs, 2015) . Today's energy revolution is a renewable one, with strong declines in renewable electricity generation costs and an observed electrification across all economic sectors, also fostered by decarbonization demands.

Electricity currently represents a share of about 20% of total energy consumption (IEA, 2020), but it is expected to become the primary energy carrier for many sectors that were previously supplied by other means, such as transportation (McCollum et al., 2014; MIT Energy Initiative, 2010), heavy industry (Lechtenböhmer et al., 2016) and – via heat pumps – home heating (Rosenow and Gibb, 2022). Due to low-cost zero-carbon technologies, such as wind and solar (Clarke et al, 2022; IEA, 2021), electricity presents the highest potential for energy decarbonization and is thus set to dominate the energy market.

Solar photovoltaics (PV), in particular, is becoming the lowest-cost electricity generation technology in ever more regions of the world (IEA, 2020; IRENA, 2019; Lazard, 2020a), with a learning rate (cost decline for each doubling of installed capacity) of 20-24% in the last decades (IEA, 2014; Wirth, 2017) and with further cost declines expected in the coming years (Fraunhofer ISE, 2015; Vartiainen et al., 2020). Beyond thus newly arising market design issues for equitable and efficient decentralized PV panel installation (Feger et al., 2022), allocative efficiency and fixed system cost recovery (Labandeira et al., 2021), increasing PV coverage is progressively challenging due to its variable nature (Baker et al., 2013;

Gowrisankaran et al., 2016). PV generation is variable at several levels, and thus the value it provides to the system differs from conventional dispatchable technologies (Hirth et al., 2016; Lamont, 2008). PV generation is seasonal in any location off-equator due to Earth's inclined angle and revolution around the Sun, entailing higher generation in summer than in winter for any given level of installed capacity. Subseasonal PV variability may be further decomposed into a diurnal cycle due to Earth rotation and intermittency due to weather variability.

Subseasonal variability can be tackled through the combination of PV generation and electricity storage capacities, as explained by Grossmann et al. (2015), and has become economically feasible thanks to the declining cost of batteries (Lazard, 2020b). However, seasonality poses a more significant challenge because seasonal energy storage is not yet an economically viable option at the required scale (Schill, 2020). PV seasonality may be addressed by either installing overcapacity to meet winter demand by PV, even at the expense of having excess production in summer; or by trading electricity – directly or transformed (e.g. possibly through hydrogen) – between locations with opposite seasonal patterns of production, significantly reducing aggregate installed capacity.

We advance the literature by combining and extending two strands of research. First, trade theory has explained the emergence of international trade since the industrial revolution, with absolute advantage (Smith, 1776) and relative (or comparative) advantage (Ricardo (1817)) describing the reasons for countries to trade and considered founding elements of economics. Heckscher-Ohlin (1991) explains trade specialization derived from differences in the relative abundance of factors of production across countries, Copeland and Taylor (1994) assess the environmental impact of North-South trade, and new trade theory developments (Bernard et al., 2007; Grossman and Helpman, 1991; Krugman, 1991, 1980; Melitz, 2003) elaborate on the subtleties of modern international trade and global value chains. Empirical applications of these models explain a substantial fraction of global trade flows in goods and services (see, e.g. Balassa, 1963; Costinot and Donalson, 2012; Davis and Weinstein, 2001; Trefler, 1995). Due to its particular physical characteristics, electricity has historically been a good not widely traded across countries at large distances. However, with the need for decarbonization and the emergence of new renewable energy technologies, in the spirit of Lancaster's (1971) demand theory, it is the time and space attributes of electricity and its availability that become ever more crucial. Correspondingly, deepening the analysis beyond undifferentiated "electricity" – which still dominates most economic evaluation to date, e.g. when relying upon standard Levelized Cost of Electricity (LCOE) – is a core requirement. When we here integrate these attributes and differentiate in the temporal (seasonal) and spatial (geographic latitude and longitude) dimension, we find that international long-distance trade in electricity can bring substantial benefits that require economic evaluation. By explicitly considering these dimensions, we identify a new source of benefits to trade, one of particularly large potential in a world entering a large-scale transition of its energy system to renewables.

Second, with the rising relevance of electricity in the context of preventing dangerous climate change, modeling the transition to a decarbonized electricity system has become critical. The economics of variable renewables differs from that of conventional dispatchable technologies, both in terms of their costs (Borenstein, 2012; Joskow, 2011) and the value they provide to the system (Gowrisankaran et al., 2016; Hirth, 2013; Lamont, 2008). This causes distortions in traditional electricity markets, such as the merit-order effect (decline of electricity prices as variable renewables penetration increase, see, e.g. Antweiler and Muesgens, 2021; Sensfuß et al., 2008), the cannibalization effect (decreasing value of variable renewables as their penetration increase, see, e.g. Hirth, 2015; López Prol et al., 2020) and increasing integration costs (Hirth et al., 2015; Reichenberg et al., 2018; Ueckerdt et al., 2013) that entail additional storage needs to complement the expansion of variable renewables (López Prol and Schill, 2021). Thus, the entry of variable renewables poses multiple challenges for the allocation of costs and benefits and the optimal policies to achieve decarbonization in the presence of externalities (Ambec and Crampes, 2019; Green and Léautier, 2015). Additionally, whereas cross-border trade in electricity has been shown to provide substantial benefits (Antweiler, 2016; Cicala, 2022), it becomes even more relevant in the context of decarbonization and large-scale integration of variable renewables (Bahar and Sauvage, 2013; Yang, 2022).

We combine and extend these two strands of research. First, by building a simple model of the electricity market in autarky comparing a variable renewable energy technology (PV), coupled with sub-seasonal storage, with a conventional dispatchable technology. Second, by extending this model to international trade, intially between two symmetric locations in opposite hemispheres in a first stage, and finally to global trade in electricity.

We then evaluate our model with current empirical data, showing that global trade in electricity could provide electricity at a cost below 22 USD/MWh plus transmission cost. Current electricity prices commonly range between 50 and 100 USD/MWh (interquartile range according to (Rademaekers et al., 2018)). Since the willingness to pay for transmission cost is thus significant and increases with the distance from the equator, as does transmission cost, we find it is likely that most locations have a willingness to pay higher than the transmission cost. In this case, there will be strong incentives for global trade of PV electricity as the i.e. cost-minimizing electricity generation option, even without acknowledging external costs of conventional electricity generation technologies, which would further increase the willingness to pay for a global grid.

To the best of our knowledge, this is the first attempt to develop a theoretical economic model of global trade in electricity and empirically evaluate it with current data. The analytical model reveals the main relationships between technologies in autarky and how they change with different trade configurations, and the empirical application sheds light on the willingness to pay for electricity trade. This paper opens fertile avenues of research regarding additional political economy considerations beyond cost minimization, such as the distribution of costs and benefits, the relocation of production capacities or the market design to enhance such large-scale coordination. Geographic characteristics with relevance for the future energy market are found to add to those identified earlier to explain the development of worldwide spatial distribution of economic activity (Henderson et al., 2018) . Likewise, this research has practical policy implications as an increasing number of country governments (notably China) and private enterprises (originating particularly from Australia and the US) are pushing strategic energy policy and/or developing business cases to exploit these very benefits of global electricity trade.

## 2. Model

This section introduces the analytical framework used to assess the advantage of global trade in electricity. First, we define the main elements of our model, followed by the derivation of the optimal combination of technologies to minimize the cost of electricity in autarky. We then elaborate a model of interhemispheric and global trade in electricity to finally assess the gains from trade and the willingness to pay for transmission, completing the analytical framework used as the basis for the empirical evaluation performed in section 3.

## 2.1. The electricity market

The perfectly homogeneous good electricity $E_{t,r,i}$ is produced in two seasons: the half-year around December (i.e. mid-September to mid-March), and June (subscript $t = \{D, J\}$); in two regions; the Northern and Southern hemispheres (subscript $r = \{N, S\}$), by two technologies: photovoltaics (PV) or baseload/backup (subscript $i = \{P, B\}$). We call the latter technology baseload when it produces a constant load of electricity, and backup when it is complementing PV, but both of them refer to the same hypothetical dispatchable conventional technology.

The cost of producing a given amount of electricity with the baseload/backup technology ($C_B$) is assumed to be a linear combination of fixed ($f_B$) and variable ($v_B$) costs and is independent of the season ($t$) and region ($r$): $C_B = f_B + v_B$ . The cost of subseasonally dispatchable PV, on the contrary, depends on the season and region and comprises only fixed costs. We assume that PV is not dispatchable across seasons, but it is dispatchable within each season. To achieve this, PV is coupled with electricity storage, so the fixed cost of subseasonally dispatchable PV is a function $h$ of fixed generation ($f_{gen}$) and fixed storage ($f_{sto}$) costs: $C_{t,r,P} = h(f_{gen}, f_{sto})$.

The two main differences between baseload/backup technology and PV are that (i) baseload/backup has positive variable costs whereas the variable cost of PV is 0, and (ii) baseload/backup technology is dispatchable, both within and across seasons, whereas PV is dispatchable only within (due to the addition of battery storage) but not across seasons. PV variability can be thus decomposed into two levels: seasonal (i.e. PV produces more in summer than in winter in any region off-equator due to Earth's inclined angle and revolution around the Sun), and subseasonal (i.e. PV experiences a diurnal cycle due to Earth's rotation and intermittency derived from weather variability). The subseasonal variability of PV can be tackled by short/medium-term electricity storage. The combination of PV generation (*G*) and storage (*S*) capacities can be optimized for any location in order to reach the minimum possible unit cost of electricity by means of G/S isolines (Grossmann et al., 2015), such that the cost of providing a given amount of electricity by subseasonally dispatchable PV is a function of the optimal fixed costs of the respective generation and storage capacities. Note, that variable costs of storage (e.g. charge losses, decline in charge efficiency) in such optimization modeling translate to a larger storage capacity demand and thus are reflected in the model as higher storage fixed costs. We assume the optimization of generation and storage capacities as fixed for any particular region under autarky, which will be further elaborated upon when assessing the gains from trade in electricity across regions.

Due to the seasonality of PV, given any level of installed capacity in a region sufficiently off-equator, the production in summer will be higher than in winter. We define the *winter hole* characteristic of PV for each region by the factor $w_r$ at which PV generation capacity needs to be added at this location to serve full winter demand (initially taken to be equal to summer demand) just by extra PV capacity:

$$w_r = \frac{\overline{E_{t,r,P}} - \underline{\underline{E_{t,r,P}}}}{\underline{E_{t,r,P}}}$$

Where the under/overline represents winter/summer PV electricity generation.

Equivalently, $a$ is the *adjustment factor* at which backup capacity is required to serve demand during the winter hole of PV:

$$a_r = \frac{\overline{E_{t,r,P}} - \underline{\underline{E_{t,r,P}}}}{\overline{E_{t,r,P}}}$$

For instance, for the latitude of Washington State, the minimum production from PV in winter is only 1/6 of its summer production. To supply a given level of PV summer production during the whole year, either 5/6 of backup capacity would have to be maintained for each unit of PV capacity (i.e. $a$ = 5/6), or 5 times more PV capacity would have to be added (i.e. w = 5), with the corresponding excess electricity generated in summer. The relationship between $a$ and $w$ is therefore given by $a_r = \frac{w_r}{w_r+1}$.

For ease of exposition, we focus on the production or supply side of electricity and assume a price-inelastic demand, initially fixed across regions and seasons (an assumption relaxed in later sections). Likewise, as lifetimes of both technologies are strongly exceeding one year, respective capacities can be assumed to be installed and available at identical levels across the within-year seasonal periods *J* and *D*. Thus, for the autarky case, it is sufficient to consider a single (within-region) electricity good, not differentiated across supply seasons. Accordingly, $\sum_i Q_{r,i}$ is the quantity of electricity consumed in region $r$ within a year, and *PV coverage* is the share $\beta_r$ of electricity consumed that has been produced by PV: $\beta_r = Q_{r,P}/\sum_i Q_{r,i}$.

PV produces more in summer than in winter in any region off-equator. The *excess threshold* ($e_r$) is the PV coverage level below which all PV electricity production meets demand, but above which excess electricity is generated during at least some period of the year. In other words, the excess threshold is the PV coverage level in overall electricity at which maximum (summer) PV production exactly meets total demand. Using a linear approximation, the excess threshold is represented by $e_r = \frac{w_r+2}{2(w_r+1)}$ (see Appendix A.1 for details). In the following, we define unit costs of electricity and analyze the cost-optimal mix under autarky before extending to interhemispheric and global trade.

### 2.2. Optimal PV coverage and unit cost of electricity in autarky

In the regions off-equator, the cost of electricity is fundamentally determined by the coverage of PV. We will first derive the cost equations depending on PV coverage, to then optimize PV coverage such as to minimize electricity cost. When PV coverage is below the point at which excess PV electricity is produced (i.e. $0 \leq \beta_r \leq e_r$, panel B of Fig. 1), the cost of electricity is the sum of three elements: (i) the cost of PV weighted by its coverage with respect to excess threshold, (ii) the cost of the backup required to adjust for the winter hole of PV, and (iii) the cost of the baseload required to cover the gap between the combination of PV plus backup and the demand level.

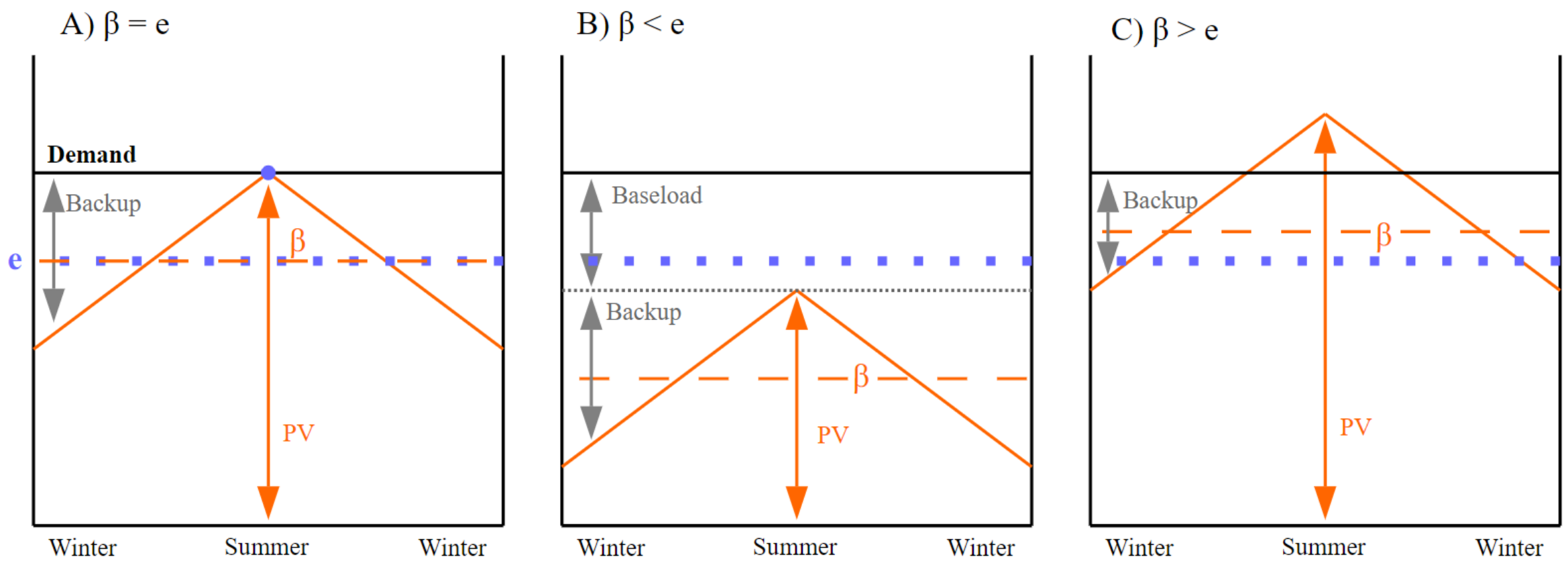

**Figure 1. Components of the electricity mix, depending on PV coverage.** All panels illustrate the modeled mix of PV generation (solid orange line) and baseload/backup (area between PV generation and the constant demand). Panel A illustrates the case where PV coverage exactly equals the excess threshold, and panels B and C where PV coverage is respectively less or greater than the excess threshold. The dashed orange line represents PV coverage (β) and the dotted blue line represents the excess threshold (*e*).

When PV coverage is higher than the excess threshold (i.e. $e_r \leq \beta_r$, panel C of Fig. 1), the cost of electricity is the sum of two elements: (i) the cost of PV, including the additional cost of the excess capacity, and (ii) the cost of the backup needed to adjust for the winter hole of PV, but only relevant for some fraction of the year.

The system of equations (1) indicates the cost of electricity in autarky in a given region ($C_r^A$) depending on whether PV coverage ($\beta_r$) is above or below the excess threshold ($e_r$), and Figure 1 shows the two general cases described above (panels B and C), and the specific case when PV coverage is equal to the excess threshold (i.e. $\beta_r = e_r$), in which case the two equations of the system (1) converge to $f_{r,P} + a_r\left(f_B + \frac{v_B}{2}\right)$. The cost of electricity in this case is the cost of PV plus the necessary backup to adjust for the winter hole of PV. When only baseload produces, the cost of electricity is simply the cost of baseload. When only PV produces, the cost of electricity is the cost of PV multiplied by one plus the winter hole coefficient to account for the necessary excess capacity.

$$C_r^A = \begin{cases} \frac{\beta_r}{e_r}\left[f_{r,P} + a_r\left(f_B + \frac{v_B}{2}\right)\right] + \left(\frac{e_r - \beta_r}{e_r}\right)(f_B + v_B) & for\ \beta_r \leq e_r \\ f_{r,P}\left(1 + \frac{\beta_r - e_r}{1 - e_r}\, w_r\right) + a_r\left(1 - \frac{\beta_r - e_r}{1 - e_r}\right) f_b + a_r\left(1 - \frac{\beta_r - e_r}{1 - e_r}\right)^2 \frac{v_B}{2} & for\ e_r \leq \beta_r \end{cases}$$

$$with\ \beta_r, e_r \in [0,1] \quad (1)$$

As $C_r^A$ is the cost of producing a given amount of electricity in a specific region in autarky with any combination of PV and baseload/backup during a year, the unit cost of electricity is simply the total cost of electricity divided by the total amount of electricity consumed ($c_r^A = C_r^A / \sum_i Q_{r,i}$). When calculating the unit cost of PV, it is important to divide total costs of PV production by the amount of PV electricity effectively consumed (i.e. excluding curtailment) to capture the cost of the excess capacity.

The higher the winter hole coefficient, the lower the excess threshold. This is illustrated in Figure 2 and occurs because the farther away from the equator (and therefore the higher the winter hole coefficient), the larger is the difference between summer and winter electricity production and therefore the lower is the PV coverage at which summer production exactly meets demand. In Figure 2 both locations “B” and “C” have the same PV coverage. However, since the winter hole coefficient is higher in location “B” (i.e. location “B” is farther from the equator), this level of PV coverage is equal to the excess threshold (because summer production equals demand). In contrast, in location “C”, with a lower winter whole coefficient, the same PV coverage is below the excess threshold (i.e. even summer production does not match total demand). Location “A” shows another situation with an even higher winter hole coefficient (even farther away from the equator) and lower PV coverage $\beta_A$ due to excess PV production.

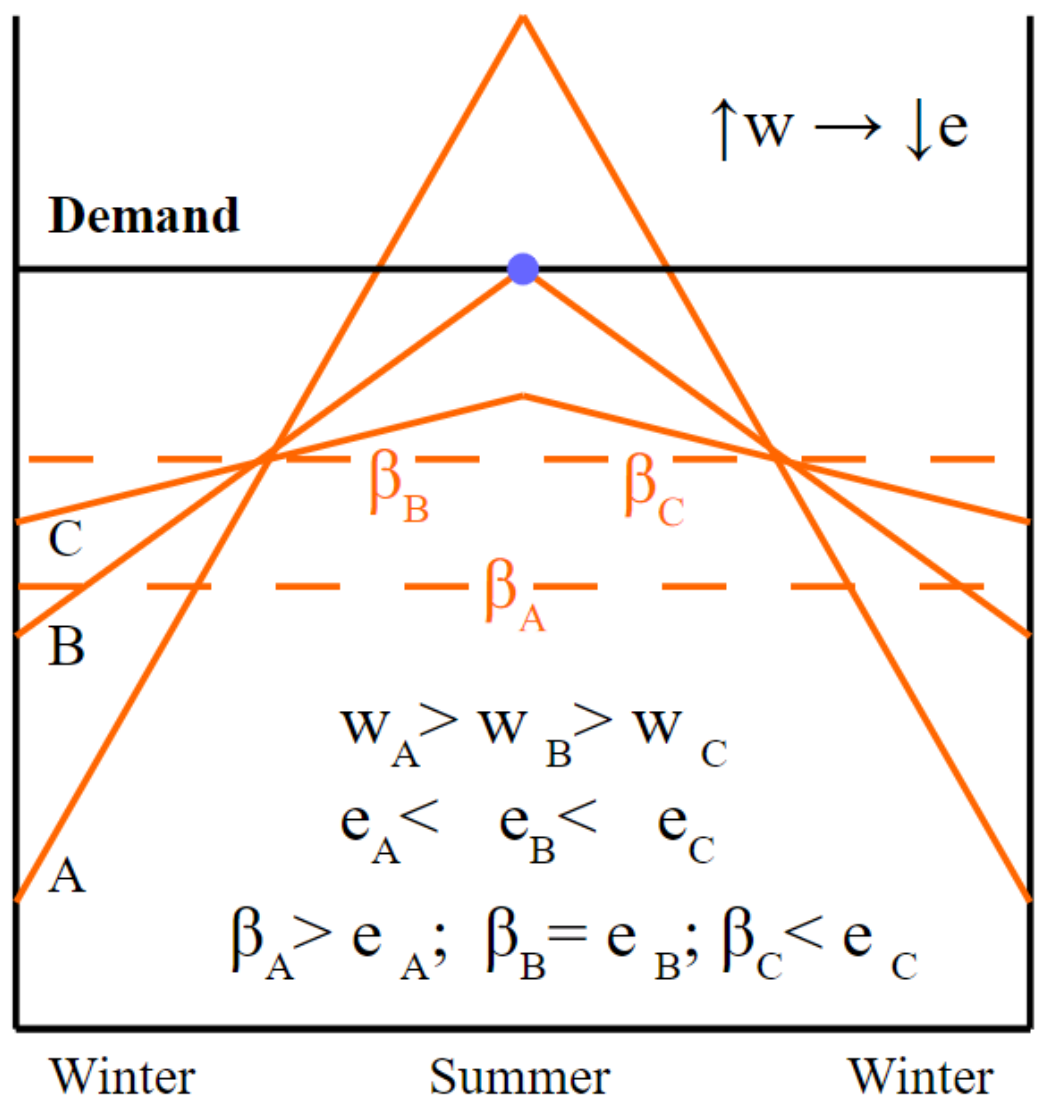

**Figure 2. Relationship between PV coverage, excess threshold and winter hole.** Lines indicated by capital letters (A, B, C) depict PV electricity production at different locations in a hemisphere over the course of a year, where each subsequent location is closer to the equator (e.g. A is farther away from the equator than B, which is farther away than C), illustrating the relationship between winter hole coefficient *w* and excess threshold *e*.

We can now derive the cost-optimal PV coverage ($\boldsymbol{\beta_r^*}$) by minimizing the cost of electricity in autarky for the two cases according to the system of equations (1):

**Proposition 1.** When $\mathbf{0} \leq \boldsymbol{\beta_r} \leq \boldsymbol{e_r}$ the optimal (i.e. cost-minimizing) PV coverage ($\beta_r^*$) is either $e_r$, when the unit costs of PV plus the necessary backup to adjust for the winter hole is lower than the unit cost of the baseload technology, or 0 otherwise.

**Proof.**

$$\min_{\beta_r} C_r(\beta_r) = \min_{\beta_r} \left[\frac{\beta_r}{e_r}\left[f_{r,P} + a_r\left(f_B + \frac{v_B}{2}\right)\right] + \left(\frac{e_r - \beta_r}{e_r}\right)(f_B + v_B)\right]$$

$$FOC: \frac{\partial C_r}{\partial \beta_r} = f_{r,P} + a_r\left(f_B + \frac{v_B}{2}\right) - (f_B + v_B) = 0$$

$$f_{r,P} + a_r\left(f_B + \frac{v_B}{2}\right) < (f_B + v_B) \Rightarrow \beta_r^* = e_r$$

$$f_{r,P} + a_r\left(f_B + \frac{v_B}{2}\right) > (f_B + v_B) \Rightarrow \beta_r^* = 0$$

Rearranging the FOC and substituting $f_{r,P}$ for $C_{r,P}$, $f_B$ for $kC_B$ and $v_B$ for $(1-k)C_B$, such that $k$ represents the share of baseload/backup fixed costs over the total baseload/backup costs, we obtain:

$$f_{r,P} = \frac{f_B}{w_r + 1} + \frac{v_B(w_r + 2)}{2(w_r + 1)} = \frac{kC_B}{w_r + 1} + \frac{(1-k)C_B(w_r + 2)}{2(w_r + 1)} = C_{r,P}$$

$$\frac{C_{r,P}}{C_B} = \frac{w_r + 2 - kw_r}{2(w_r + 1)} \Leftrightarrow w_r = \frac{2(1 - C_{r,P}/C_B)}{2\,C_{r,P}/C_B - 1 + k} \tag{2}$$

Equation (2) is illustrated in Panel A of Figure 3, where the surface represents the threshold between *0* and $e_r$ PV coverage. This threshold is determined by the winter hole coefficient (vertical axis, $w_r$), the share of fixed costs over the total baseload/backup costs (left horizontal axis, *k*), and the relative cost of PV with respect to baseload/backup (right horizontal axis, $C_{r,P}/C_B$). The area above the plane represents an optimal PV coverage of *0*, whereas the area below the plane represents an optimal PV coverage of (at least) $e_r$. Figure 2 Panel A could be read as "what is the minimum relative cost of PV at which PV coverage enters the market (in which case it reaches a coverage equal to the excess threshold $e_r$), given the winter hole coefficient in a specific location and the cost structure of the baseload/backup technology".

The cost of electricity at the equator (i.e. $a = w = 0$)[b] is directly determined by the one technology (PV or baseload) that supplies electricity at the lowest cost (see proof in Appendix A). At the equator, therefore, only the cheaper technology enters the market; thus the straight line along $C_{r,P}/C_B$ equal to 1 independently of the cost structure of the baseload/backup technology. As the winter hole increases, PV needs to become cheaper with respect to the baseload/backup technology (i.e. the relative price of PV with respect to baseload/backup ($C_{r,P}/C_B$) has to decline) to enter the market. Since the backup technology supports PV coverage, and backup is cheaper with increasing shares of variable cost (because the cost structure of backup is $f_B + \frac{v_B}{2}$ whereas the cost structure of baseload is $f_B + v_B$), the following holds: the higher the share of fixed cost of the baseload/backup technology (*k)*, the cheaper PV has to get to enter the market. Since both *w* and *k* are inversely related to $C_{r,P}/C_B$, the shape of the surface expands as it moves away from the $w_r k$ vertex (i.e. the farther away from the $w_r k$ origin of coordinates, the cheaper (lower $C_{r,P}/C_B$) PV has to become relative to baseload/backup to enter the market). This reflects the fact that PV is better complemented by peaking backup (high variable, low fixed cost) than by baseload (opposite cost structure) technologies.

[b] Note that the assumption a = w = 0 is an approximation to the actual situation, which is more subtle due to the obliquity of the ecliptic. Our numerical evaluations showed that not only at the equator but up to 8° N and S the results from this approximation are sufficiently similar to the precise value; the impact of attenuation from weather is much higher.

Likewise, we can obtain the optimal PV coverage for the range of PV coverage between the excess threshold and 1 in the same way.

**Proposition 2.** When $\boldsymbol{e_r \leq \beta_r \leq 1}$ the optimal PV coverage is given by equation (3):

$$\beta_r^* = \frac{a_r v_B + a_r f_B (1 - e_r) - w_r f_P\,(1 - e)}{a_r v_B} \tag{3}$$

**Proof.**

$$\min_{\beta_r} C(\beta_r) = \min_{\beta_r}\left[f_{r,P}\left(1 + \frac{\beta_r - e_r}{1 - e_r}\,w_r\right) + a_r\left(1 - \frac{\beta_r - e_r}{1 - e_r}\right) f_B + a_r\left(1 - \frac{\beta_r - e_r}{1 - e_r}\right)^2 \frac{v_B}{2}\right]$$

$$FOC\colon \frac{\partial C_r}{\partial \beta_r} = \; f_{r,P}\,w_r - a_r f_B - a_r v_B\left(1 - \frac{\beta_r - e_r}{1 - e_r}\right) = 0$$

$$\beta_r^* = \frac{a_r v_B + a_r f_B (1 - e_r) - w_r f_P\,(1 - e_r)}{a_r v_B}$$

The second derivative is positive and therefore the FOC represents a local minimum:

$$\frac{\partial^2 C_r}{\partial \beta_r^2} = \; a_r v_B \left(\frac{1}{1 - e_r}\right)^2$$

We can express eq. (3) as a function of $k$ and $C_{r,P}/C_B$ by substituting $f_{r,P}$ for $C_{r,P}$, $f_B$ for $kC_B$ and $v_b$ for $(1-k)C_B$ in the FOC as in Proposition 1. We also substitute $a_r$ and $e_r$ for their respective values such that:

$$f_{r,P}\,w_r - a_r f_B - a_r v_B\left(1 - \frac{\beta_r^* - e_r}{1 - e_r}\right) =$$

$$C_{r,P} w_r - \frac{w_r}{w_r + 1} k C_B - \frac{w_r}{w_r + 1}(1 - k) C_B \left(1 - \frac{\beta_r^* - \frac{w_r + 2}{2(w_r + 1)}}{1 - \frac{w_r + 2}{2(w_r + 1)}}\right) = 0$$

Isolating $\beta_r^*$ and rearranging we obtain eq. (4):

$$\beta_r^* = \frac{C_{r,P} w_r^2 + 2kC_B + \left(kC_B - 2C_B + C_{r,P}\right) w_r - 2C_B}{2(kC_B + (kC_B - C_B) w_r - C_B)} =$$

$$\frac{C_{r,P}}{C_B} \quad \frac{w_r^2 + w_r}{2(k + kw_r - w_r - 1)} + \frac{2(k - w_r - 1) + kw_r}{2(k + kw_r - w_r - 1)} \tag{4}$$

Equation (4) allows us to illustrate the optimal PV coverage ($\beta_r^*$) as a function of the share of fixed cost over total baseload/backup cost ($k$) and the ratio between PV and baseload/backup cost ($C_{r,P}/C_B$) for any given region (i.e. its respective winter hole coefficient). Figure 3 Panel B shows the optimal PV coverage according to equation (4) for two levels of the winter hole coefficient (0.5, corresponding to e.g. north-west Australia, and 10, corresponding to e.g. northern Germany), already combined with the threshold given by equation (2) and illustrated in Panel A of the same figure. The lines along $k$ and $C_{r,P}/C_B$ corresponding to $w = 0.5$ and $w = 10$ in Panel A become the vertical planes in Panel B, representing the threshold at which PV coverage goes from 0 to $e_r$ in accordance to proposition 1.

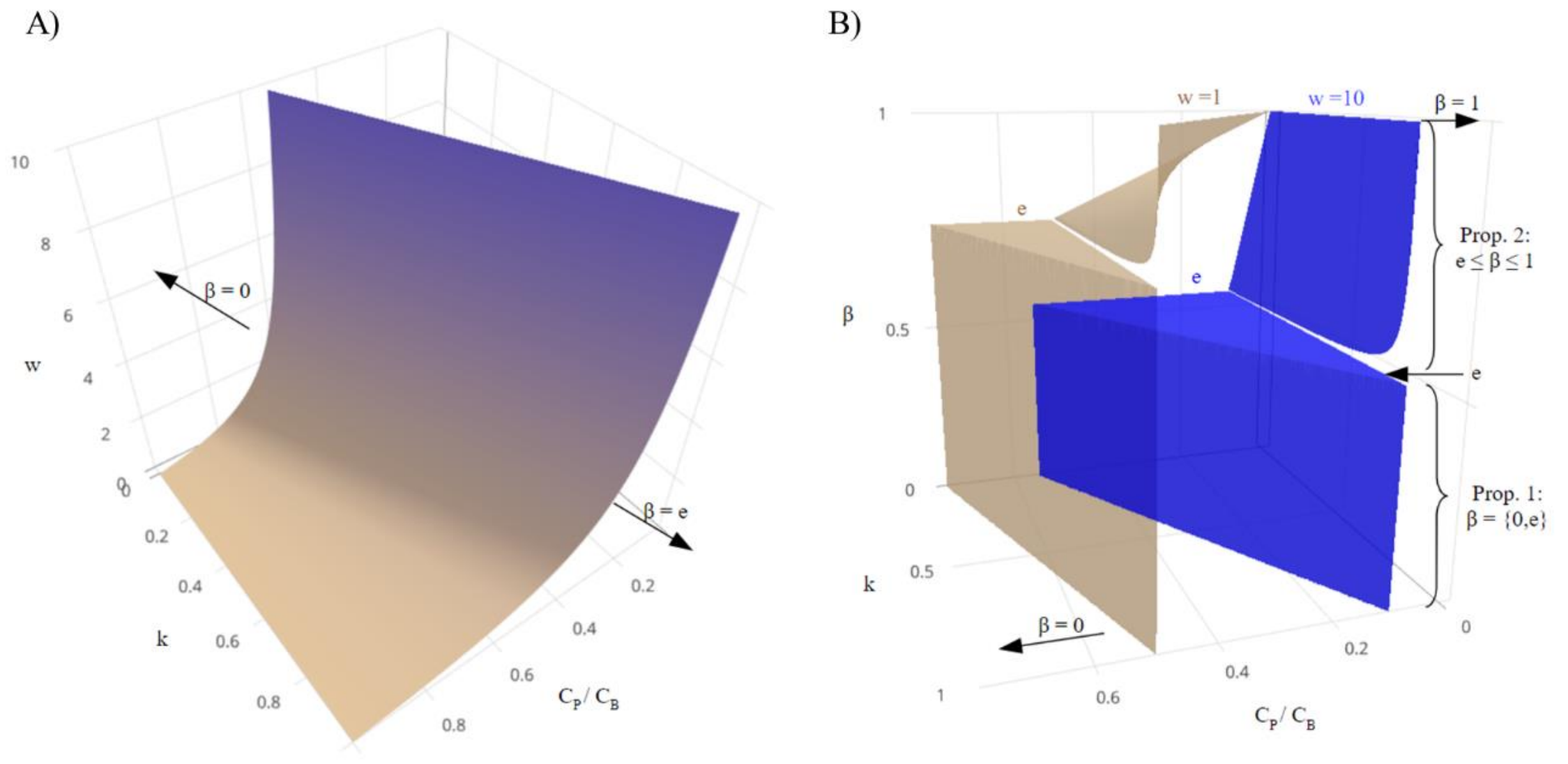

**Figure 3**. **Illustration of Proposition 1 (left), and propositions 1 and 2 combined (right).** The surface in panel A depicts the threshold at which PV coverage switches from zero to equaling the excess threshold *e*, as described by Equation (2), determined by the winter hole

coefficient $w$, share of fixed costs in total baseload/backup costs $k$, and the ratio of PV to baseload/backup costs $(C_{r,P}/C_B)$. Panel B depicts optimal PV coverage for winter hole coefficient levels of 0.5 (light brown) and 10 (dark blue). An interactive version of the figure can be found at: https://kawilliges.shinyapps.io/PV-global-trade/. This interactive version is available for readers individual sensitivity analysis, including for specific latitudes and respective levels of the winter hole coefficient.

As mentioned in the previous section and illustrated theoretically in Figure 2, the higher the winter hole coefficient, the lower the excess threshold. That is why the vertical plane of the low winter hole surface (w = 0.5, light brown) is higher than that of the high winter hole surface (w = 10, dark blue). The horizontal plane of each surface corresponds to their respective excess threshold, from which it is increasingly difficult to expand PV coverage due to the increasing cost of the excess capacity.

The excess threshold reverses the relationship between the share of fixed cost over total baseload/backup cost ($k$) and the optimal PV coverage ($\beta_r^*$). Whereas at PV coverage levels below the excess threshold backup supports PV coverage (i.e. the cheaper the backup, the higher PV coverage; Figure 1 panel B), when PV coverage is above the excess threshold, backup does not support, but rather competes against PV coverage (i.e. the cheaper the backup, the lower PV coverage; Figure 1 panel C).

The $k$ $C_{r,P}/C_B$ space to the left of each of the planes represents the area in which PV coverage is 0 in each of the locations because PV is not sufficiently cheap to compensate for the cost of accommodating the winter hole. On the opposite, once the surface reaches 1, the $k$ $C_{r,P}/C_B$ space to the right represents a 100% PV coverage, i.e. all electricity is produced by PV.

Figure 3 allows us to understand the relationship between variables and how the baseload/backup cost structure effect on PV coverage reverses at the excess threshold. It also illustrates the relevance of the winter hole for achieving higher PV coverage levels and the increasingly difficult task of increasing PV coverage above the excess threshold. The interactive version of Figure 3 available for readers at https://kawilliges.shinyapps.io/PV-global-trade/ can be applied for individual sensitivity analysis, including for specific latitudes and respective levels of the winter hole coefficient.

### 2.3. Interhemispheric and global trade in electricity

Once we have derived the optimal coverage of PV in autarky, we can expand our analysis to interhemispheric and global trade of electricity. We will first develop a model of bilateral North-South (N-S) and East-West (E-W) interhemispheric trade in electricity; the N-S model will then be generalized to $n$ regions to produce a model of global electricity trade.

For the analysis of bilateral N-S electricity trade we assume that both regions are perfectly symmetric in terms of demand, solar insolation (i.e. $E_{J,N,P} = E_{D,S,P}$ and $E_{D,N,P} = E_{J,S,P}$) and trade (or transmission) cost and have equal technological costs and investment endowment. We also assume that unit cost of subseasonally dispatchable PV is lower than the unit cost of baseload/backup (otherwise baseload would be the only generating technology in any situation).

Figure 4 shows the production/consumption possibility frontier of both regions in autarky and trade, where the vertical axis ($E_J$) denotes half-year electricity around June and the horizontal axis ($E_D$) denotes half-year electricity around December. Since both regions are symmetrically off-equator, PV produces more in each region's summer than in winter with opposite seasonal patterns ($A_{r,P}$). Since baseload/backup is seasonally dispatchable, resources could be devoted to higher capacity (and fixed costs) of baseload/backup, but operation for only a fraction of the year, and thus it could produce more in one season than in the other, yet with higher unit cost. Only equal production in both seasons entails 100% utilization of the installed capacity (year-round) and therefore the lowest unit electricity cost. The latter point is represented by ($A_B$) and the backup/baseload production possibility frontier is represented by the concave curve crossing that point. The line ($\overline{A_B A_{r,P}}$) represents any possible combination between PV and baseload producing constantly across seasons for either region.

Assuming constant demand across seasons (i.e. 45º line departing from the origin), both regions' autarky optimum would be to produce only with baseload ($A_B$). If a region produced only with PV while having a constant demand year-round, its maximum demand level would be the winter PV production, while the difference between summer and winter production would be excess electricity. This is equivalent to building PV overcapacity to serve the winter hole.

When both regions trade, and assuming no trade cost, constant demand across seasons and symmetric trade across regions (i.e. both regions export and import the same inducing a net zero balance of trade over the year), the optimum would be for both regions to produce only with PV and to trade electricity to balance out their opposite seasonal patterns (reaching point $T$ in Figure 4. This would increase total production and consumption, yielding gains from trade equivalent to the distance between the autarky and the trade points ($\overline{A_B T}$).

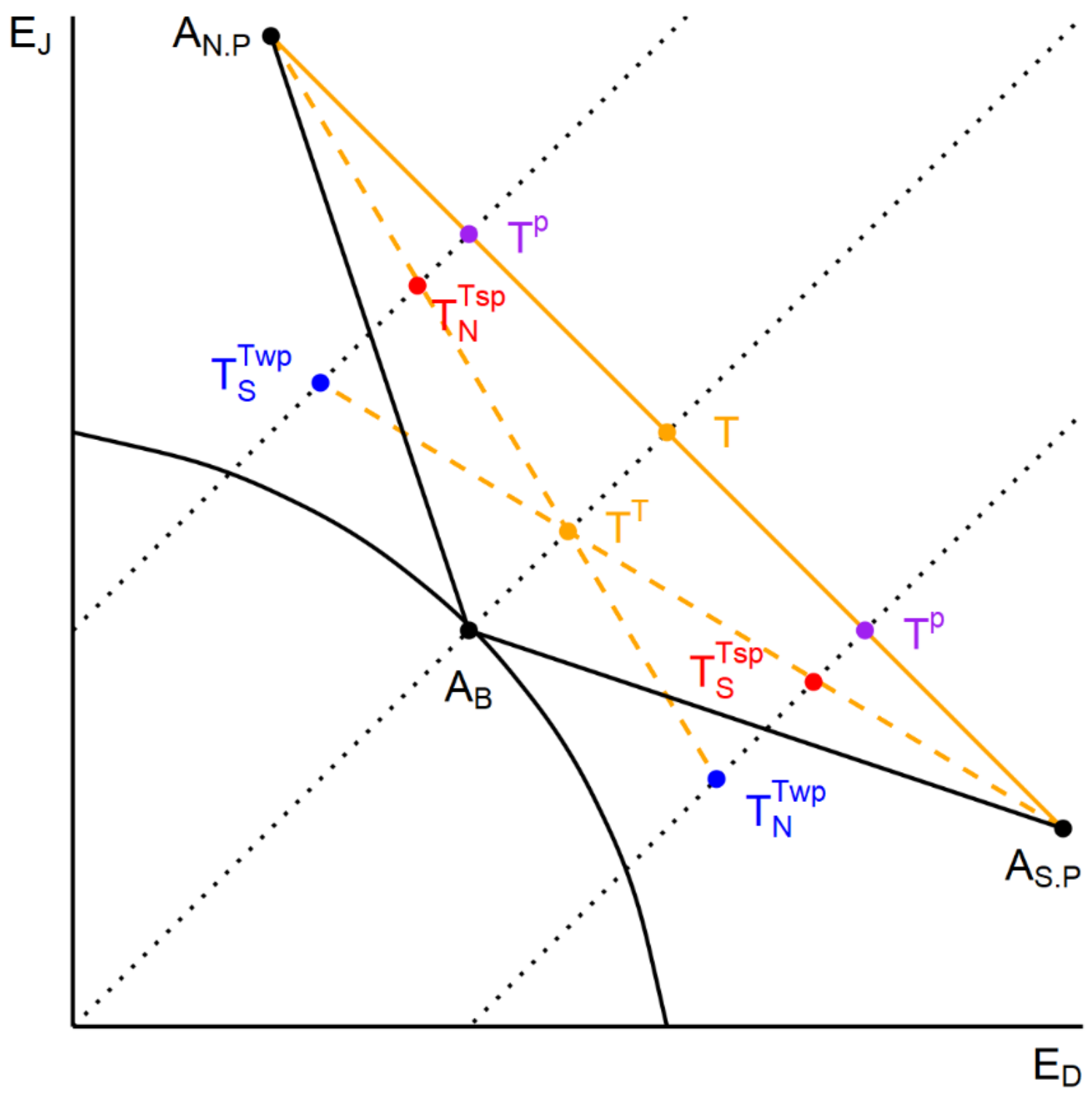

**Figure 4. Interhemispheric trade in electricity.** The figure depicts the production/consumption possibility frontier in both autarky (black lines) and interhemispheric trade (orange lines). The dashed orange line indicates trade with the inclusion of trade costs, with the red and blue points indicating higher demand in summer and winter, respectively.

The constant demand across seasons assumption can be relaxed such that both regions consume more in one season than in the other. In this situation, both countries would be symmetrically (as far as there is an equalized balance of trade) situated along the trade line (solid orange), e.g. as the pair of purple points ($T^p$).

Trade or transmission costs can be represented by rotation of the trade line departing from each region's PV autarky situation and diverging from the no-trade-cost trade line more with increasing trade volume as we move away from the autarky point. This line (dotted orange in Figure 4) represents the increasing transmission capacity costs as the consumption pattern of a region differs from its original PV seasonality. Including trade costs, the trade point becomes $T^T$ instead of $T$ and therefore the gains from trade are reduced from $(\overline{A_B T})$ to $(\overline{A_B T^T})$, with $(\overline{T^T T})$ therefore representing trade cost.

If demand differs across seasons, whether it is a winter or a summer peaking electricity system does not matter if we ignore trade cost, but it matters when we consider them. As the consumption pattern differs from PV seasonality, the trade equilibrium moves "down" (in the sense of lower equilibrium total production/consumption possibilities, i.e. closer to the origin of coordinates) along the trade line to a lower level equilibrium due to higher unit trade costs. This is represented by the higher-level equilibrium of the summer peaking regions $(T_r^{Tsp})$ relative to the winter peaking $(T_r^{Twp})$ ones. Whereas the equilibrium of the summer peaking trade is above the autarky production possibility frontier and therefore trading pays off, the winter peaking trade equilibrium is below the autarky production possibility frontier combining some PV and some backup. This entails that a combination of both technologies in autarky would be more efficient than only PV with trade due to high trade costs associated to a consumption pattern opposite to PV seasonality. Likewise, if demand differs across regions, the utilization rate of the transmission line declines, increasing its unit cost and reducing the potential gains from trade.

The E-W trade is equivalent to the N-S with the difference that the variability offset by trade is the diurnal rather than the seasonal cycle of N-S trade. The necessary amount of generation and storage capacity is endogenously adjusted accordingly, as further explained in Section 2.4. Since the diurnal cycle is easier to solve with short/medium-term electricity storage, in our empirical section below the gains from trade turn out generally lower than those from N-S trade.

We can generalize the bilateral N-S trade to *n* regions. Figure 5 shows an empirically based yet stylized cross-section of the Earth from the north pole (dot on the vertical axis) to the south pole (dot on the horizontal axis), passing through the tropics and the equator, where the solid black line depicts PV production for every location in autarky. The "M" shape of this line is derived from the empirical observation that the maximum PV potential occurs along the tropics rather than at the equator (see Figure 7, panel C), due to aspects such as a larger share of air water vapor/cloudy weather in the latter.

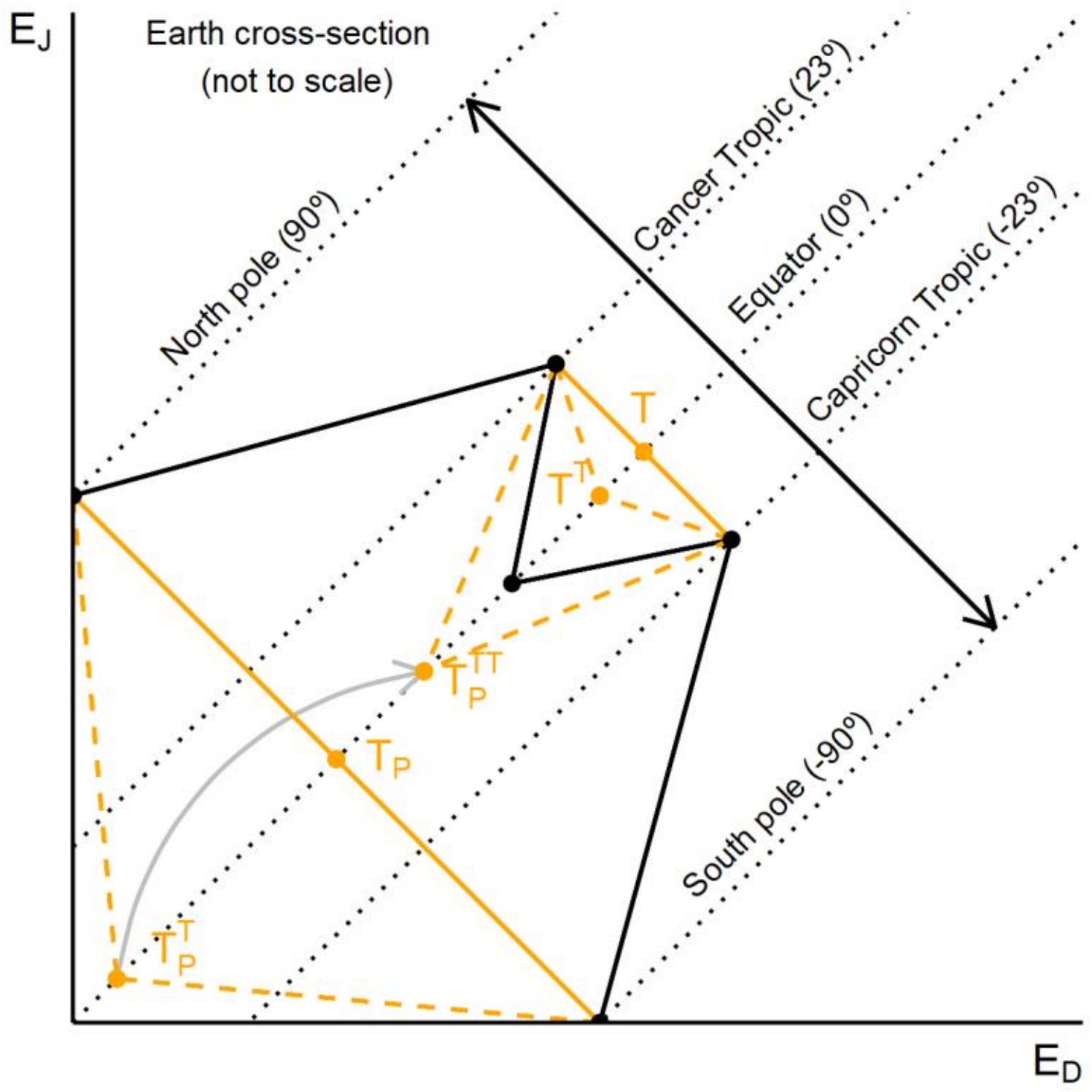

**Figure 5. Global trade in electricity.** The figure depicts a cross-section of the Earth with the North pole as a point on the vertical axis, and the South pole on the horizontal axis. The solid black line indicates PV production at each latitude in autarky, with costless trade drawn in solid

orange, and trade with transmission costs depicted as dashed orange lines. As an example of the benefits of the global grid, the grey arrow represents the gains of locating PV capacity in the optimal locations (the tropics) and then importing the electricity generated to the consumption centers (in our extreme case, the poles).

Equivalent to what we have seen for the bilateral case, the tropics could trade in order to reach a constant production across seasons reaching $T$ without trade cost or $T^T$ when trade costs are included. The same holds for the extreme cases of the poles, which could reach constant production across seasons by trading between them reaching $T_P$ or $T_P^T$ when trade costs are included. In this case, trade costs are much higher than in the case of the tropics, given both the larger difference with respect to their respective PV seasonal patterns and the larger distance between both locations. Ignoring any other political economy considerations, the poles could reach a higher level of consumption by allocating PV capacity to the best Earth locations on the tropics and importing all the electricity from there. Assuming the same trade costs, relocating capacity to the best locations would increase the consumption possibilities of the poles in the amount $\overline{T_P^T T_P^{TT}}$, represented by a grey arrow in Figure 5.

## 2.4. Gains from trade and willingness to pay for transmission cost

The unit gains from trade ($G_r$) are simply the difference between the unit cost of electricity in autarky and the unit cost of electricity with trade, minus unit transmission cost $TC_r$ (eq. (5)). We can therefore define the willingness to pay for transmission cost ($WTP_r$) as the difference between autarky and trade unit cost of electricity (eq. (6)).

$$G_r = c_r^A - (c_r^T + TC_r) \tag{5}$$

$$WTP_r = c_r^A - c_r^T \tag{6}$$

Figure 6 illustrates the three paradigmatic situations in autarky and off-equator: (i) production with only baseload ($\beta = 0$), (ii) production with PV plus backup ($\beta = e$), and (iii) production with only PV (i.e. $\beta = 1$) with the excess capacity and curtailment according to the respective winter hole coefficient. Figure 6 also shows the components of trade gains: (i) N-S trade removes the costs caused by seasonality (either backup, excess capacity or a combination of both); (ii) E-W trade removes the costs caused by the diurnal cycle, and global trade removes both of them, but also (iii) removes the additional cost caused by weather variability[c], and (iv) reduces the unit generation cost by allocating capacity in the best locations with highest solar irradiation.

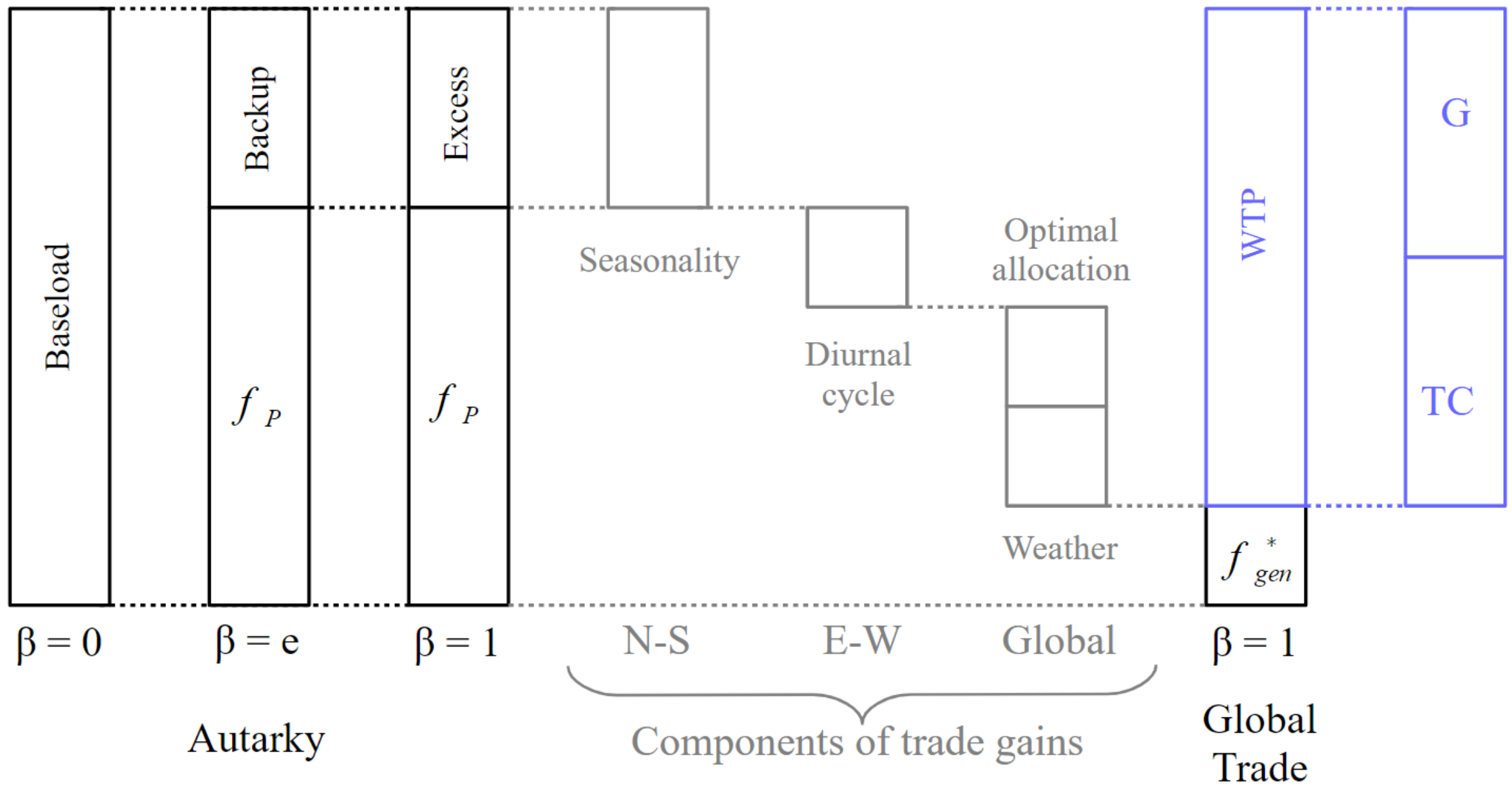

**Figure 6. Components of electricity costs and trade gains.**

[c] We assume that only a global grid scenario would address short-term variability caused by weather, and do not consider interhemispheric trade (either N-S or E-W) – in this case representing trade between just two locations – able to address such issues with balancing resources.

The type of trade regime (autarky, North-South, East-West or global) determines the unit cost of subseasonally dispatchable PV. Both in autarky and in the bilateral N-S trade the unit cost of subseasonally dispatchable PV is the same because N-S trade removes seasonality (i.e. $w$ becomes 0), but it does not affect the subseasonal pattern. According to the method developed by Grossmann et al., (2015), the unit cost of subseasonally dispatchable PV is a function of the PV generation and storage capacity that minimizes the unit cost of PV electricity ($f_{r,P} = h(f_{gen}, f_{sto})$). This can be generalized by defining the *overcapacity factor* (OCF: the optimal amount of PV generation capacity given by the generation-storage isoline with respect to the minimum PV capacity that would be necessary to produce the same quantity of electricity consumed ignoring insolation variability), and the respective storage to generation capacity ratio (S/G: storage capacity per unit of generation capacity). $E_{r,P}$ is the PV electricity generated per unit of generation capacity in a specific location and $AIC_{gen}$ and $AIC_{sto}$ the annualized investment cost (including all fixed costs) of the generation and storage capacity respectively. As we assume that N-S interhemispheric trade cannot be used for subseasonal balancing purposes, OCF, S/G ratios and AICs remain the same in cost calculations for a given location between autarky and N-S trade.

In the E-W trade case, the diurnal cycle is removed, so the cost of subseasonally dispatchable PV is reduced accordingly. The minimum necessary overcapacity factor (OCF) to tackle the diurnal cycle is 1.05 (i.e. an equal amount as produced for immediate consumption has to be produced to be stored, assuming a 10% storage loss), and the minimum necessary storage capacity per unit of generation capacity ($S/G$) is half of daily generation (i.e. $E_{r,P}/730$). We can therefore derive the optimal unit cost of subseasonally dispatchable PV as the difference between the unit cost given by the generation-storage isolines and the unit cost of the minimum storage and generation capacities needed to tackle the diurnal cycle.

With global trade, both seasonal and diurnal cycle, as well as weather variability are solved by the global integration of the solar resource, which is of roughly stable influx at the global level. This means that, on the one hand, storage is no longer needed[d], so the fixed cost of PV only comprises the cost of generation capacity. On the other hand, as illustrated in Figure 5, with global trade, all capacity would be allocated to the locations with the highest solar irradiation, and therefore the fixed cost of generation capacity becomes the optimal fixed generation cost ($f_{gen}^{*}$). The unit cost of subseasonally dispatchable PV depending on the type of trade is summarized in the system of equations (7).

$$c_{r,P} = \begin{cases} h(f_{gen}, f_{sto}) = \dfrac{OCF \cdot AIC_{gen} + \frac{S}{G} \cdot AIC_{sto}}{E_{r,P}} & \textit{in autarky and } N-S \textit{ trade} \\ h(f_{gen}, f_{sto}) = \dfrac{(OCF - 1.05)}{E_{r,P}} AIC_{gen} + \dfrac{730\frac{S}{G} - 1}{730\, E_{r,P}} AIC_{sto} & \textit{in } E-W \textit{ trade} \\ f_{gen}^{*} = \dfrac{AIC_{gen}}{E_{r,P}^{*}} & \textit{in global trade} \end{cases} \quad (7)$$

## 3. Numerical evaluation

### 3.1. Data and assumptions

Building upon our analytical model derived in Section 2, we now estimate the willingness to pay for interhemispheric and global interconnections through the use of empirical data. We obtain (i) the unit cost of electricity, (ii) the optimal PV coverage and (iii) the willingness to pay for transmission cost in each of the trade cases in all Earth locations between ±55º latitude. Trade increases welfare when the unit transmission cost is lower than the willingness to pay. We use real-world data on current generation costs of conventional and PV technologies and 20 years of historical high-resolution solar insolation data as described below.

[d] In practice, some storage would still be needed for balancing purposes. This storage capacity is however negligible with respect to the magnitude of the system so we omit it.

To build a representative baseload/backup technology, we take the world production shares of the three main conventional electricity generation technologies: coal, natural gas and nuclear (53.2%, 32.6% and 14.2% respectively (BP, 2019), and further disaggregate between combined cycle and gas peaking according to US generation data (Dubin, 2019). Finally, we take the costs for each technology (high range without externalities according to Lazard, 2019) and weigh them by their respective shares to produce the unit cost of a synthetic baseload/backup technology. For the scenario described in this section, we assume this synthetic technology has annualized fixed costs of 779 thousand USD per year per megawatt, and variable costs (in unit terms) of 23.3 USD per megawatt-hour. Alternatives to this relatively high-cost main scenario are discussed in section 4.2 and can be found in more detail together with an elaborated description of the calculation of the synthetic technology costs in Appendix B.

To calculate the unit cost of subseasonally dispatchable PV (based on Eq. 7) for every location, OCF and S/G ratio values are required, which are calculated based on interpolation of calculated values for 31 land locations at varying latitudes, according to the method developed by Grossmann et al. (2015). We assume a cost of 500 USD/kWp and a lifetime of 30 years for PV generation capacity and 200 USD/kWh and 15 years respectively for battery storage capacity. This can be considered a conservative assumption, given the latest and future expected technological and cost developments (IRENA, 2019; Lazard, 2020a; Vartiainen et al., 2020).

The isolines approach of Grossmann et al. (2015) allows for the determination of the optimal mix of generation and storage capacities at any given location. We calculate isolines for a moving average period of 50 days, empirically found to be the necessary timeframe to ensure subseasonally dispatchable PV generation. A moving average allows for removal of the day-night cycle and intermittency due to shorter-term instances of cloud cover and allows for isolating the effect of the winter hole on PV production (see Figure B.2 in Appendix B for a graphical depiction of this isolating effect). Panels A and B of Figure 7 depict the results of the isolines optimization and subsequent interpolation of OCF and S/G ratio, respectively, for every degree of latitude.

The remaining component of unit cost calculations from Eq. (7) is PV yield per unit of installed capacity, obtained from a database of long-term yearly average of daily totals of potential PV electricity production (Solargis, 2017, Figure 7, Panel C). The highest potential yearly PV production occurs around the tropics of Cancer and Capricorn (around 23 degrees N and S), rather than the equator, due to differences in direct and diffuse sunlight (driven by dust, pollution and water vapor in the atmosphere), day length and solar altitude (Page, 2012).

The winter hole coefficient, necessary for calculating optimal PV penetration as described in system of equations 1 from section 2 is calculated as the 50-day maximum global horizontal irradiation minus the minimum, divided by the minimum for a representative year, which is constructed from average daily horizontal irradiation data for 20 years, according to data obtained from the NASA Langley Research Center Atmospheric Science Data Center Surface meteorological and Solar Energy web portal supported by the NASA LaRC POWER Project (NASA, 2019). Results of this calculation are found in panel D of Figure 7.

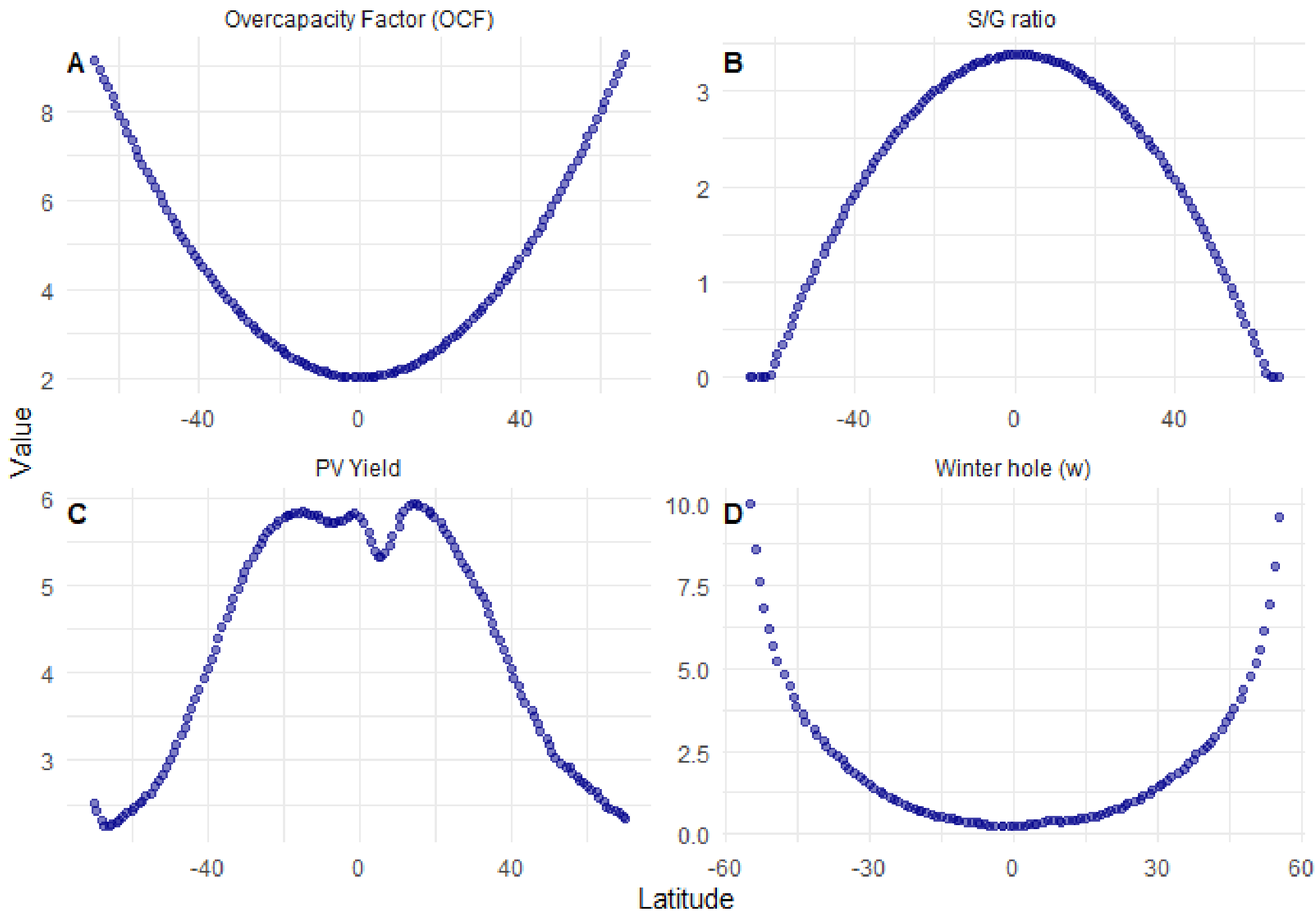

**Figure 7. Summary of model input data by latitude.** Panel A depicts the overcapacity factor (OCF) of PV to handle subseasonal – including diurnal cycle and weather – variability and Panel B the storage to generation capacity ratio, both derived via isoline interpolation. Panel C shows modelled yearly average PV yield, and panel D the size of the winter hole coefficient.

While winter hole coefficients increase exponentially at the poles (approaching the limits of the Arctic/Antarctic circles, where due to periods of no light at all, the coefficient is undefined), the majority of the global population lives between 55 degrees North and 15 degrees south; the highest winter hole coefficient in such areas is below 10. We thus limit our empirical evaluation to latitudes between 55 degrees North and South.

## 3.2. Results

### 3.2.1. Optimal PV coverage and unit cost of electricity in autarky

The data and assumptions described in Section 3.1 allow for calculation of (i) unit cost of subsesonally dispatchable PV (according to equation 7), (ii) optimal PV coverage (according to equation 3), and (iii) the unit cost of electricity (according to the system of equations 1). Results for the autarky case – using costs of a synthetic baseload/backup as described in Appendix B – are shown in Figure 8, and as latitude-specific averages in Figure B.3 (in the Appendix).

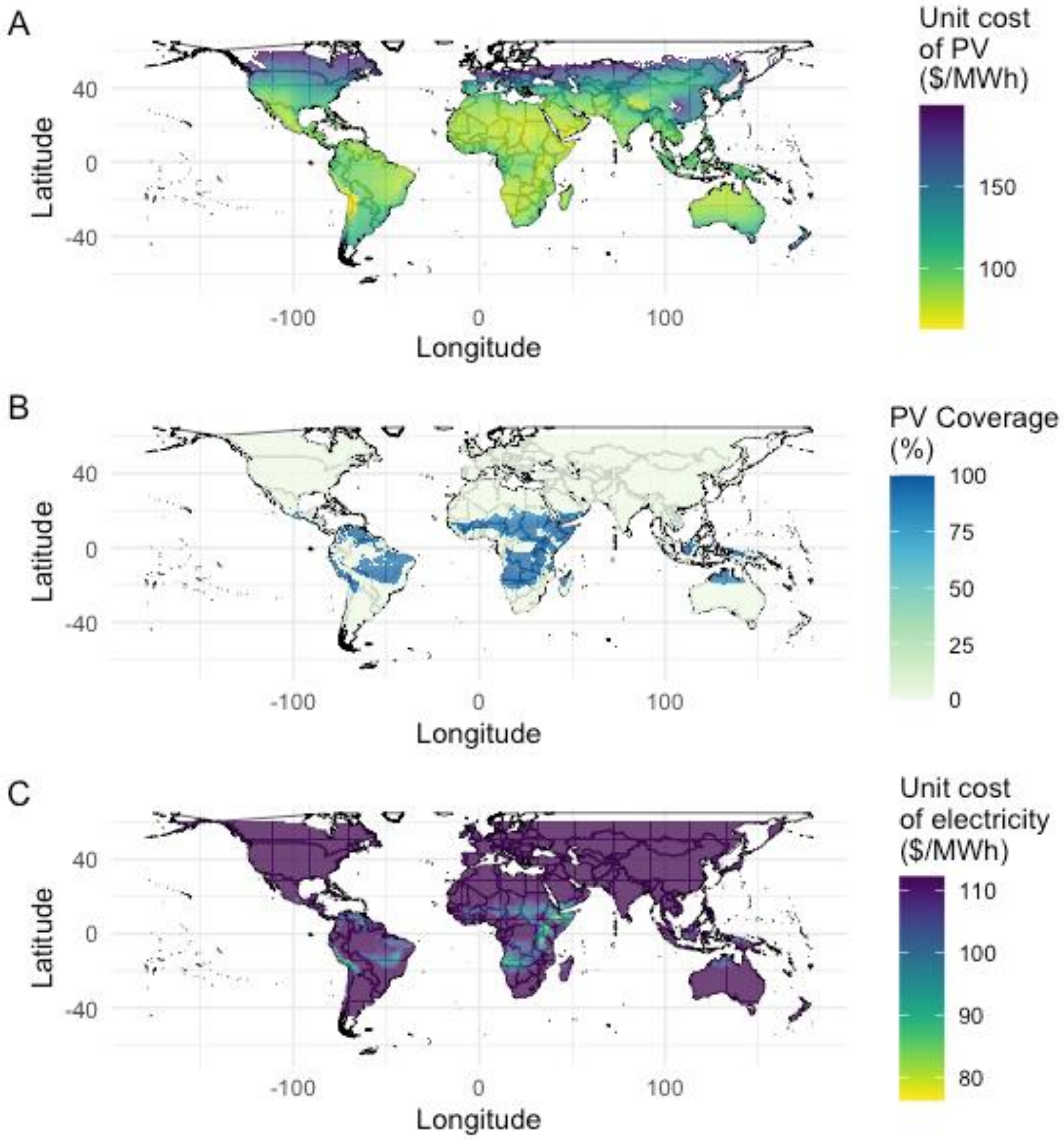

**Figure 8. Empirical model results for the autarky case.** Panel A shows the modelled cost of subseasonally dispatchable PV, panel B the PV coverage in autarky, and panel C the unit cost of electricity in autarky, using fixed and variable cost estimates for the synthetic baseload/backup mix as described in Appendix B and unit cost of PV as described in equation (7).

As shown in panel B of Figure 8, PV coverage is negatively correlated with the location-specific winter hole coefficient such that the higher the winter hole coefficient (i.e. the farther away from the equator), the lower is the cost-optimal PV coverage. While for many locations, the unit cost of PV is below the cost of a system using only the baseload/backup technology (e.g. the South American west coast, south of 20 degrees latitude) PV coverage is still equal to zero due to the cost of accommodating the winter hole with more expensive backup generation. In areas where PV enters the market, PV coverage $\beta$ is equal to the excess threshold $e$ (of note is that in autarky, PV coverage never entirely dominates the electricity mix, the maximum is around 97%, as seasonal overcapacity and curtailment is too expensive). The sharp transition from areas with high PV coverage to zero penetration coincides with rising values of winter hole coefficient and corresponding rapid decline in potential PV yield outside of the tropics. As the winter hole increases, the excess threshold decreases. Thus, in autarky, PV only penetrates in areas with a low winter hole coefficient, and those regions also show relatively high levels of $e$, determining the penetration rate.

Determination of optimal PV coverage allows for the calculation of the unit cost of electricity in autarky according to equation (1). The results are shown in the lower panel of Figure 8. PV only enters the market in equatorial regions where the winter hole is low and therefore accommodating it with backup is not very costly (panel B). Where PV enters the market, the unit cost of electricity in autarky declines.

### 3.2.2. Unit cost of electricity with trade and willingness to pay for transmission

Figure 9 shows the unit cost of electricity (as determined by the system of equations 1) and willingness to pay for transmission (calculated as unit costs in autarky minus the cost in trade) under the three trade scenarios described in Section 2.4.

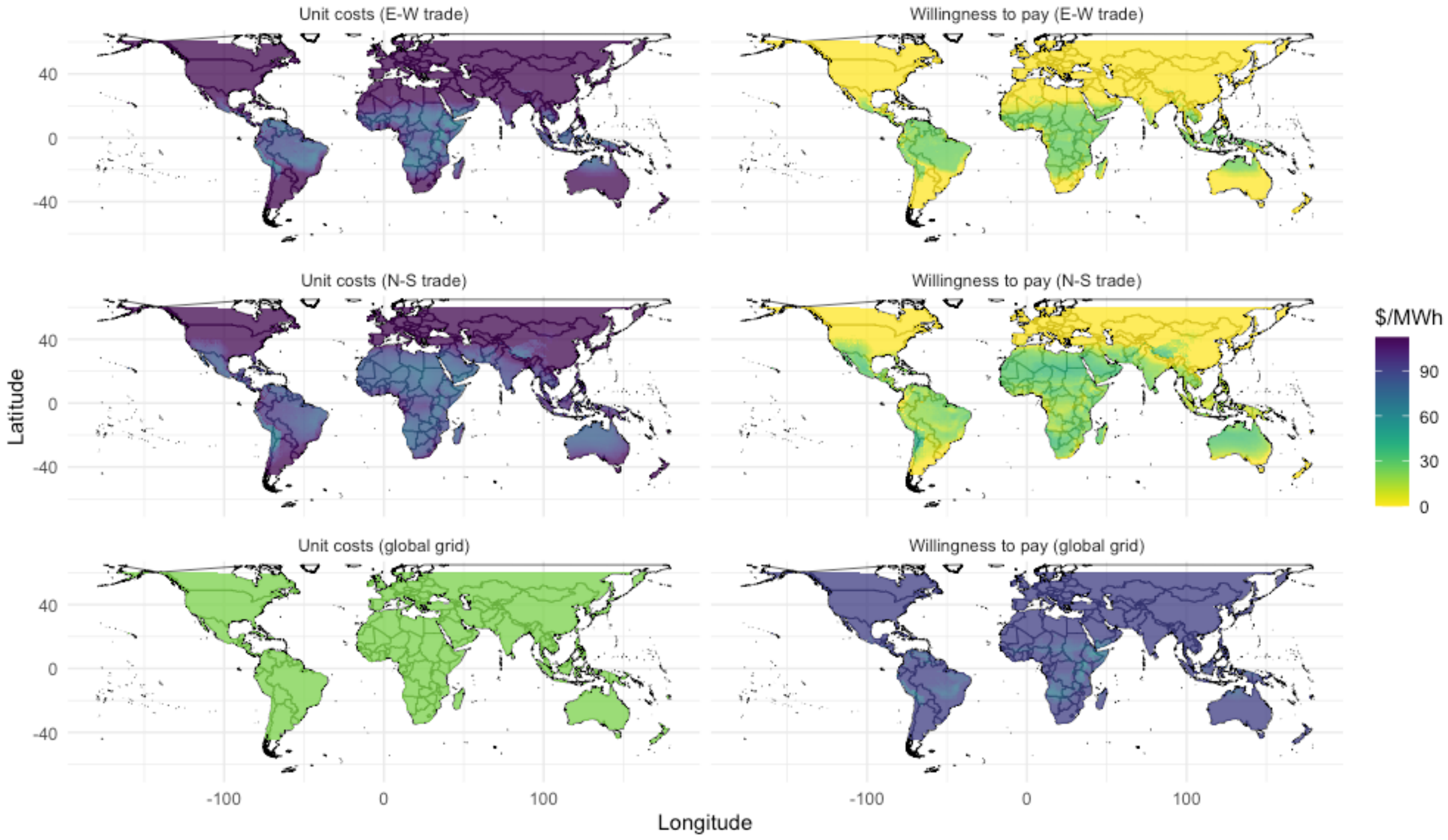

**Figure 9. Unit costs of electricity and willingness to pay for electricity trade.** Unit cost of electricity (left column) and willingness to pay for transmission (right column) under East-West (top), North-South (middle) and global grid (bottom) trade scenarios. An online tool depicting these results as well as allowing for interactive sensitivity analysis can be found at: https://kawilliges.shinyapps.io/PV-global-trade/.

The ability for trade across longitudes, e.g. east-west transmission, is depicted in the top row of Figure 9. As we concluded in Section 2.4, the reduced need for overcapacity and minimum storage capacity reduces unit costs of PV and subsequently increases optimal penetration, lowering the unit cost compared to the autarky case, particularly in regions closer to the equator, where seasonality is not a limiting factor to penetration in the autarky case, in comparison to the day-night production limitation.

The North-South trade regime (middle row of Figure 9) demonstrates the strong effects of offset seasonal patterns compensating for each other, thereby removing the winter hole (and thus resulting in changing unit costs of electricity as defined by Equation (1), given that trade in electricity effectively reduces winter hole and adjustment factor parameters (*w* and *a)* for PV to zero). The situation is analogous to considering all locations on the globe to be situated at the equator – seasonality does not exist – and penetration in each location is determined by the lowest unit cost technology. In other words, if the cost of PV provided by Equation (7) is less than the cost of synthetic baseload/backup, PV penetration equals 1, but does not enter the market otherwise. As shown in Figure 9, a North-South trade regime results in higher PV penetration and lower unit costs for locations much farther away from the Equator than East-West trade. Downward effects on unit costs and increased willingness to pay for trade are stronger as a result of North-South trade as compared to East-West for latitudes beyond 15 degrees off the equator. However, for locations within ±15º, East-West trade provides higher gains than North-South due to the lower relevance of seasonality compared to the diurnal cycle (Figure B.3 in the Appendix shows the magnitude of individual effects at varying latitudes).

The bottom row of Figure 9 depicts the unit cost of electricity and willingness to pay for a global grid. Such a grid would merge the benefits of both East-West and North-South interconnections, i.e. offsetting both diurnal and seasonal patterns (respectively). Additionally, a global grid would allow for the allocation of PV capacity in optimal locations on Earth (as shown stylistically in Figure 5) and removes the need for storage (as shown in Figure 6). As a result, the unit cost of PV in a global grid (from equation 7) is only the fixed cost of generation in optimal locations, which is lower than the unit cost of PV in any other circumstance. For our global grid scenario, we assume that the optimal location allows for a long-term average annual

PV production of 1,515 kWh/kWp (the first quartile value from Solargis data on potential PV electricity production (Solargis, 2017), which would represent that global installed PV capacity is spread uniformly within the best half of locations).

Additional results for alternative cost scenarios, as well as analysis of the sensitivity of results to varying costs of baseload/backup, can be found in Appendix B.2, and an online interactive tool is provided for further investigation at https://kawilliges.shinyapps.io/PV-global-trade/.

Our model, evaluated with current data, suggests that while subseasonally-dispatchable PV would not be able to compete with a baseload/backup generation technology (based on real-world energy mixes and costs) in autarky beyond the excess threshold, introducing global trade would imply a willingness to pay for traded PV electricity to allow for 100% PV coverage at the low cost provided by the best Earth locations (the exact cost depending on the spread of the installed capacity across the best locations). East-West trade causes even regions with some PV in autarky to have substantial willingness to pay due to elimination of the day-night cycle, while North-South trade would allow for PV penetration into a much larger range of latitudes. In a global grid electricity system, PV would produce electricity at a cost of 21.47 USD/MWh, plus transmission costs, resulting in willingness to pay for trade for the majority of the globe of above 90 USD/MWh. In the next section, we discuss the implications of our model findings on the need for a global renewable energy transition and address limitations and potential extensions to our work.

## 4. Discussion

### 4.1. International trade enhancing welfare in the global renewable energy transition

The complementary advantage observed in the bilateral interhemispheric electricity trade implies that, assuming symmetric trade and consumption, both regions increase their production/consumption thanks to specialization. They do not only specialize in one good (the one in which they have absolute advantage), but also in one technology, since baseload/backup is displaced by PV in both regions.

Global trade in electricity entails complementary advantage at the global level in the sense that every region can increase their consumption possibilities by relocating capacity to the best regions and importing the electricity produced. There is, however, intrahemispheric absolute advantage in the sense that all production is concentrated in the best locations of each hemisphere. This poses a substantial challenge in terms of political acceptability. However, provided the significant potential benefits from trade, and the fact that solar insolation is widely spread across the globe, it might be possible to find a mutually beneficial and stable equilibrium from this type of international cooperation, as shown by Churkin et al. (2019) for the case of East Asia.

### 4.2. Modelling abstractions

We have developed an abstract idealization of a stylized electricity system that allows us to understand the dynamics of variability and the opportunities of interhemispheric and global trade to obtain the maximum possible value of the lowest-cost electricity generation technology: solar photovoltaics. Our intention is not to describe current deployment trends, but to explore the cost-optimal pathway to decarbonize electricity supply. For this reason, our results differ from reality. PV coverage in our model is lower than in reality off-equator for three main reasons: (i) the cost of sub-seasonally dispatchable PV includes in our model the storage required to tackle sub-seasonal variability, whereas in reality this is usually done by already existing dispatchable capacity; (ii) PV has higher value at low penetration due to its correlation with demand, but we assume constant demand (i.e. a completely flat demand profile); and (iii) we omit support policies and internalization of external costs. On the other hand, our model provides higher PV coverage within the tropics than observed in reality, as we ignore the political economy considerations that hamper PV diffusion in these areas.

We also assume a single cost of baseload/backup, PV, and storage capacities in both the analytical and numerical evaluation, resulting in a single worldwide unit cost of electricity when PV does not enter the market. Using a single cost for each component allows us to focus on the effects of interhemispheric and global trade, with the trade-off of not accurately representing spatially heterogeneous costs across regions. In reality, electricity prices vary widely, with wholesale prices ranging globally between approximately 50-100 USD/MWh (representing 1st and 3rd quartile values for selected countries; the highest value reported was Brazil with an estimate of over 175 USD/MWh) (Rademaekers et al., 2018).[e] Energy costs are usually higher than wholesale prices due to them being covered partly also by capacity payments, subsidies and other remuneration mechanisms complementing wholesale markets. Thus, while the numerical evaluation cannot reflect country-specific differences in autarky, our results are in a similar range, with unit costs of electricity approaching 115 USD/MWh. Moreover, the model is flexible enough to allow for further spatial disaggregation.

---

[e] While referring to the end of 2017, this being the most recent comprehensive disaggregated data published to date, price ranges can be considered to have remained in the same order of magnitude except for the recent shocks caused by the pandemic and the subsequent recovery that are expected to be temporary.

Finally, the results presented above are based on a high-cost scenario for baseload and backup according to the increasing risks and therefore financing costs of fossil fuels. Scenarios with lower costs would reduce PV penetration and unit costs in both autarky and trade. The median cost scenario presented in Appendix II drastically reduces unit costs in autarky to just over 85 USD/MWh. However, East-West and North-South trade still results in higher PV penetration and lower unit costs (see Figure B.4 in Appendix B), while the low-cost scenario would result in a baseload/backup cost that does not allow PV to enter the market given our PV generation and storage cost assumptions. Figures B.5 and B.6 further illustrate the sensitivity of results to changing costs of baseload and backup, and the reader can build their own scenarios in the application provided by the authors at [https://kawilliges.shinyapps.io/PV-global-trade/](https://kawilliges.shinyapps.io/PV-global-trade/).

### 4.3 Potential extensions

Our model is aimed at identifying the relationship between variables and the potential benefits of interhemispheric and global trade in electricity. Keeping it simple allows us to understand these relationships, as shown in Figs. 1 and 2. To achieve this simplicity, we make some simplifications that could be relaxed in future extensions of our model, and that we summarize as follows:

(i) Technological: we assume only one renewable energy technology (PV), one conventional (baseload/backup) and one type of storage (batteries). Several renewable energy technologies could be considered (e.g. wind, geothermal, biomass). This would decrease renewables' aggregate variability and therefore result in higher renewables coverage in autarky. Since global trade removes all variability and PV is the lowest cost generation technology in the optimal locations, the global trade estimates would not change, but the willingness to pay could decrease if combining different renewable technologies can reduce the autarky unit cost. Likewise, baseload and backup technologies could be separately considered (e.g. nuclear for baseload and gas for backup), taking into account their different cost structures, and storage capacities could be endogenously optimized depending on PV penetration. Finally, as hydrogen becomes competitive, it could be considered as an alternative energy carrier rather than transmission of electricity itself, or means of storage or as a complement to PV overcapacity or backup to overcome seasonality.

(ii) Functional forms: we consider linear approximations. This is sufficient to identify relationships between variables and produce approximate estimations, but more flexible functional forms derived from empirical data would produce more accurate estimates.

(iii) Constant demand: we assume constant demand and focus on the supply side. Future research could integrate demand flexibility as a way to reduce the cost of seasonality (be it transmission cost, overcapacity or backup). This would also be useful to estimate the full welfare effects of trade. Since the generation profile could be adjusted to the long-term demand pattern in a global grid by distributing the power plants accordingly across the globe, the constant demand assumption is unlikely to have a significant effect on the unit cost estimation.

(iv) Trade in embodied electricity: relaxing the initial assumption of symmetric demand across seasons and regions implies higher transmission costs due to lower utilization rates (Section 2.3). An alternative to higher transmission cost would be to maximize transmission utilization and use the surplus electricity in the lower-demand region (the Southern hemisphere) to shift trade patterns towards production and export of electricity-intensive goods (e.g. hydrogen through electrolysis, or steel produced thereof). In this case, the gains from trade would also include the comparative advantage in these electricity-intensive goods produced with the surplus electricity in the lower-demand region.

(v) Technical and political feasibility: we have not discussed the technical and political feasibility of laying down a global electricity grid. Some of the technical challenges are related to the deployment of long-distance subsea cables and the real-time management of a global grid. The political challenges are related to the necessary international cooperation required to achieve such a global project, the distribution of costs and benefits and the willingness of countries to accept such levels of energy dependence. What our analysis shows, however, is that these issues are likely to move higher on the political (and technological) agenda, against the background of the substantial cost-advantage derived from such interhemispheric/global cooperation.

(vi) Regulation and market design: electricity systems' regulations and market design vary widely across countries. For this reason, it is difficult to imagine a global electricity market with homogeneous regulation across countries. However, there is a trend in regional integration and homogeneization across neighbouring countries and states, e.g. with the "Energy Union" in the European Union (European Commission, 2021), the progressive adoption of wholesale markets across US states (Cicala, 2022) and the creation of an unified national electricity market in China (National Development and Reform Commission, 2022). More research could compare and identify optimal design features to be adopted across regions (e.g. Cramton, 2003).

(vii) Static vs. dynamic: our model is static; it highlights the potential benefits of global trade in electricity but it does not study the process to get there. However, the different trade configurations analysed point at the possibility of initially connecting specific locations with opposite seasonal patterns that could set the examples for more regions to join at later stages. Other models could expand on this idea to have a more detailed picture about the pathway, possibly with energy/electricity system models such as in Victoria et al. (2020).

(viii) Partial vs. general equilibrium: we here focus on the effects of global trade in electricity to provide the lowest possible unit-cost electricity worldwide. This development would have spillover effects to other economic sectors within countries and trade patterns across regions, induced by both output but also input markets (the latter also including e.g. material demands for cables and batteries). For this reason, it would be interesting to further analyze equilibrium effects with macroeconomic computable general equilibrium models (e.g. Bachner et al., 2019).

### 4.4. Towards a global electricity grid

We present a static analysis of electricity trade in different scenarios (interhemispheric and global trade), but have ignored the dynamic process towards such grid integration scenarios. Whereas a global electricity grid may seem a far-fetched idea, countries and regions are progressively moving towards greater electricity market integration. The European Union has a long-term strategy for an energy union with unified market rules and an interconnection capacity of 15% of annual production (European Commission, 2017). Likewise, China and the USA are working on expanding their grids at a quasi-continental scale (Bloom et al., 2020; Fairley, 2019), and most studies assessing interconnections agree on the potential benefits of grid expansion at continental levels and beyond (Aghahosseini et al., 2019; Grossmann et al., 2014, 2013; Jacobson, 2021).

The trade configurations we propose, while not currently in operation, are not purely theoretical. High voltage direct current (HVDC) generation has seen increased use in recent decades to provide long-distance electricity transmission, as it is much more cost-effective than traditional alternating current transmission and experiences less electricity loss over distance. While until recently, the focus of HVDC use was on regional transmission, with lengths on the order of hundreds of kilometers, the feasibility of longer distance transmission has been investigated particularly in terms of techno-economic analysis (see e.g. Ardelean et al., 2020; Ardelean and Minnebo, 2017; Purvins et al., 2018). Most recently, a cable over 3,000 kilometers in length, capable of transmitting 12 gigawatts, was completed in China (Hitachi, 2019), and additional connections are planned of similar, potentially hemisphere-spanning lengths, such as the proposed Sun Cable project to link Australia and Singapore with a cable over 4,500 km in length (Sun Cable, 2020).

These developments allow for an approximation of potential transmission costs. For a typical distance of mid-latitudes to the equator, such as the 4,500 km of the Sun Cable project, the literature based on data of the early 2010s (e.g. MacDonald et al., 2015) indicates transmission cost of about 28.5 US$/MWh (see Appendix C). Cost declines since then imply that the current transmission costs (construction and maintenance, including all overheads) for the Sun Cable project could be an estimated 19.8 USD/MWh for this distance, and costs are expected to decline to 5 USD/MWh within the current decade (for details see Appendix C). Particularly in the case of a global grid configuration as we model it, the excess of willingness to pay over electricity production cost as identified in section 3 (and depicted in Figure 8) for this distance by far exceeds respective transmission costs, and likewise for other locations sufficiently off equator. This indicates a substantial potential gain of trade in electricity, with expectations for further increases in the future.

## 5. Conclusions

We develop a theoretical model for interhemispheric and global trade in electricity. This is relevant in a world experiencing widespread electrification and seeking fully renewable energy supply, as PV becomes the lowest-cost, most widely available and most sustainable source of electricity, but with its diffusion being hampered by its variability (seasonality, diurnal cycle and short-term intermittency). Interhemispheric trade in electricity can address this variability. North-South electricity trade removes the PV seasonality, East-West trade removes the diurnal cycle, and global integration removes both patterns and virtually all intermittency, yet each at the cost of transmission. Cost minimizing electricity production in the presence of a sufficiently expanded global electricity grid entails that PV produces all the electricity consumed as soon as the unit cost of PV in the best locations is lower than the unit cost of the alternative baseload/backup technology (as it is already the case).

Conversely, we find that it is increasingly difficult for regions in autarky to ramp up PV coverage above what we define as the excess threshold. Since the excess threshold (the point at which summer PV generation exactly matches demand) declines with the absolute latitude, whereas tropical regions can reach high shares of PV coverage at low cost even in autarky, high absolute latitude regions will not be able to increase PV coverage at low cost without enhancing electricity market integration or other forms of seasonal flexibility. The analytical model also sheds light on the competition dynamics in autarky as PV penetration increases. Below the excess threshold, PV competes against baseload and benefits from low-cost back-up. However, once PV reaches the excess threshold and all baseload has been displaced by PV, low-cost backup prevents the further penetration of PV.

Since both transmission costs and the willingness to pay for it increase monotonically with latitude, it is likely possible to find a mutually beneficial and stable equilibrium to allocate the costs and benefits of a global grid. There are political challenges, however, derived from the concentration of PV capacity in the regions with the best resource, energy dependence and regulatory integration at supra-national levels. For these reasons, there may be a trade-off between political acceptability and lower unit cost. To be on the cautious side, we have used the first quartile level of potential PV production, showing that even a widespread distribution of PV capacity in the best half of Earth locations would achieve a low unit cost of electricity with a global grid. Our model suggests that such a grid would result in the majority of areas on Earth having a willingness to pay for electricity trade of over 90 USD/MWh. Whereas this could be a development opportunity for some of the world's least developed regions, it might be difficult for states to accept a situation of chronic energy dependency. On the other hand, the existence of a dense interconnected global network would minimize the risks of blackouts due to local environmental conditions, providing an additional resiliency value, particularly relevant in the context of increasing extreme events due to climate change.

Further research should explore the political economy implications of a globally integrated electricity system and the potential options for a mutually beneficial stable equilibrium that could allow such a situation as well as the macroeconomic spillover effects, changes in trade patterns and development opportunities. Our model could be likewise expanded by considering additional conventional (e.g. differentiating backup and baseload) and renewable (e.g. solar vs. wind) electricity generation and storage (e.g. hydrogen or pumped hydro) technologies, or by considering elastic demand that could serve as a flexibility option to allow higher PV coverage even in autarky.

For the global transformation to achieve the Sustainable Development Goals, and in particular the Paris Agreement climate targets, our results indicate that a successful shift of the energy system to a renewable basis can be fostered significantly by global trade to keep the costs of such a transformation low. These findings are crucial and warrant consideration in guiding the design of the currently transitioning global energy system. More generally, the significant willingness to pay identified hints at an imminent shift to be seen on the political horizon towards consideration of long-distance trade in renewable and cheap electricity. The economic stakes for its implementation are high, and imply political economy issues far beyond the energy sector. The way the international community will respond to this opportunity will shape how economic activity will tend to relocate across the hemispheres.

## Appendix A. Analytical model details

### A.1. Derivation of the excess threshold

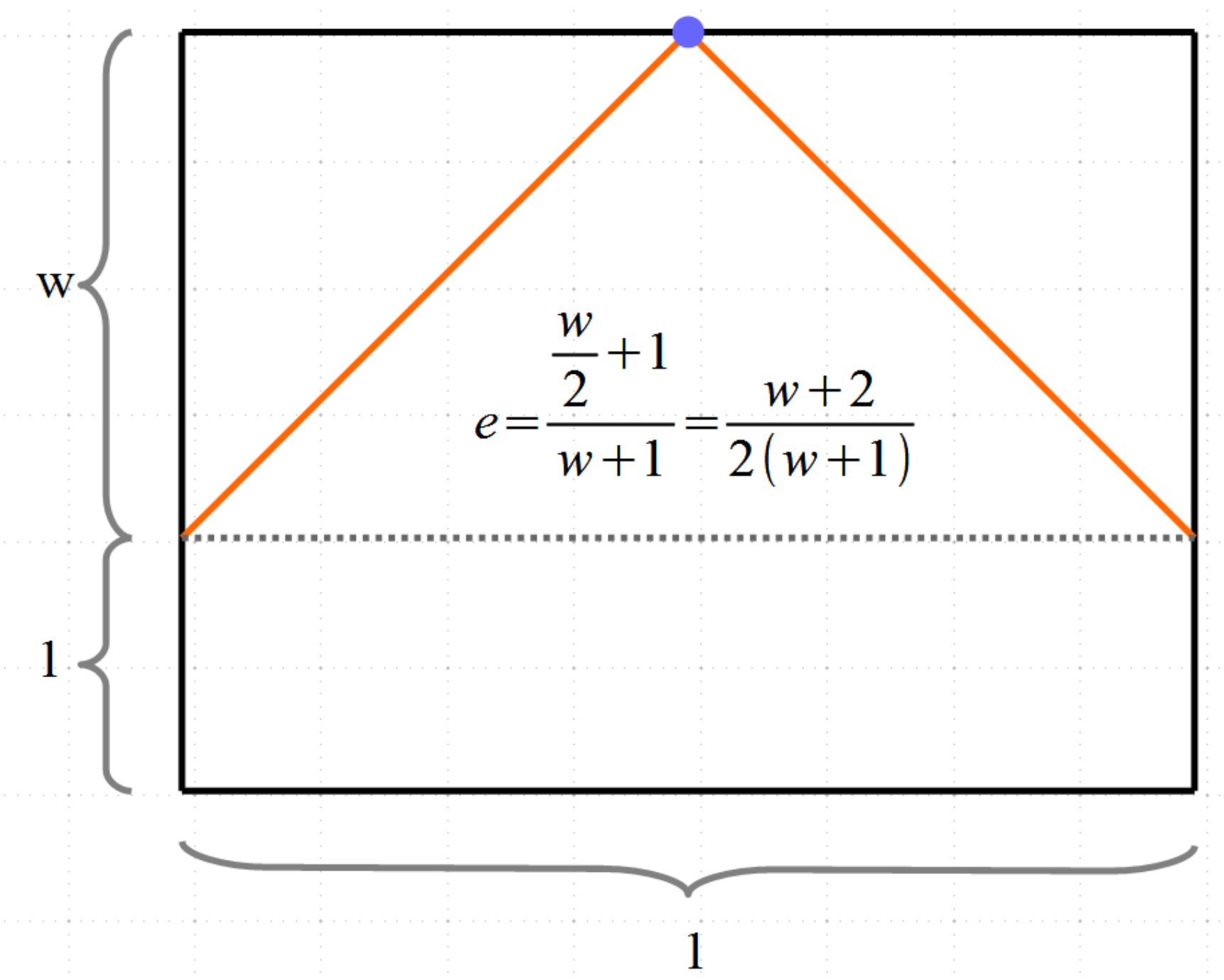

**Figure A.1. Derivation of the excess threshold**

### A.2. Proposition 3

**Proposition 3.** When $\boldsymbol{w = 0}$ the optimal PV coverage is either 0 when the cost of PV is higher than the cost of baseload or 1 otherwise

**Proof.**

$$\min_{\beta} \; C(\beta) = \min_{\beta} \left[\beta f_{pv} + (1-\beta)\,(f_b + v_b)\right]$$

$$FOC\colon \frac{\partial C}{\partial \beta} = \; f_{pv}\,(f_b + v_b) = 0$$

$$\beta^* = 1 \;\Leftrightarrow\; f_{pv} \; < (f_b + v_b)$$

$$\beta^* = 0 \;\Leftrightarrow\; f_{pv} \; > (f_b + v_b)$$

**A.3. Derivation of East-West trade:**

$$\frac{OCF * CRF_g + \frac{S}{G} * CRF_s}{E_{pv}} - \left(\frac{1.05 * CRF_g}{E_{pv}} + \frac{CRF_s}{730}\right)$$

$$= \frac{(OCF - 1.05)}{E_{pv}} CRF_g + \left(\frac{\frac{S}{G}}{E_{pv}} - \frac{1}{730}\right) CRF_s$$

**Appendix B: Further details regarding the numerical evaluation**

**B.1. Construction of a synthetic baseload/backup technology**

We begin by defining the share of natural gas, coal and nuclear power that will make up the eventual synthetic technology. As oil is mainly used in island and isolated systems, we exclude it from our baseload/backup technology. Using shares of world gross electricity generation from (BP, 2019) (p. 56), we arrive at an initial makeup of the synthetic technology of 32.6% natural gas, 53.2% coal, and 14.2% nuclear power.

Within natural gas, we have baseload (Combined cycle) and peaking technologies (combustion and steam turbines). As they differ both in costs and capacity factors, it is important to differentiate between the two. Using data from Dubin (2019), we find the capacity and generation shares of each gas technology; combined cycle plants make up 56.5% of natural gas capacity, but represent 85.7% of generation, while peaking plants make up 43.5% of capacity and 14.3% of generation.

Combining the estimated generation share of nuclear, coal and both gas technologies produces the final generation share for our representative baseload/backup, consisting of 53.2% coal, 27.91% gas combined cycle, 14.2% nuclear, and 4.65% gas peaking. Costs for each technology are derived from data collected from Lazard (2019), summarized in Table B.1, where the annualized fixed cost comprises capital costs (at 7.7% WACC as assumed by Lazard) and the fixed O&M cost and is given as USD/kW of installed capacity per year (or equivalently thousand USD/MW per year); and the variable cost comprises fuel costs (taking into account differentiated heat rates by technology as presented by Lazard), and variable O&M cost and is given in unit terms, i.e. USD/MWh (i.e. to supply a constant load of 1MW all hours of the year we would have to multiply times 8760). Technology-specific capacity factors are also derived from Lazard (2019) and can be found in Table B.1.

**Table B.1. Annualized fixed costs, unit terms variable costs, and capacity factors of dispatchable electricity generation.**

| | $\boldsymbol{f_i}$ AFC (Annualized Fixed Cost) k USD/MW per year | | | $\boldsymbol{v_b}$ in unit terms vble. Cost (USD/MWh) | | | $\boldsymbol{CF_i}$ (%) | | |
|---|---|---|---|---|---|---|---|---|---|
| | Min | max | mean | min | max | mean | min | max | mean |
| Gas peaking | 75 | 115 | 95 | 32.35 | 40.07 | 36.21 | 10 | 10 | 10 |
| Nuclear | 669 | 1123 | 896 | 12.38 | 13.13 | 12.76 | 91 | 90 | 90.5 |
| Coal | 284 | 589 | 437 | 15.44 | 22.40 | 18.92 | 83 | 66 | 74.5 |
| Gas CC | 81 | 143 | 112 | 24.16 | 27.56 | 25.86 | 70 | 55 | 62.5 |

We calculate the fixed and variable cost for the synthetic backup technology as follows:

$$f_b = \sum\left(f_i * s_i * {CF_i}^{-1}\right); \quad v_b = \sum(v_i * s_i)$$

where the variable cost is simply calculated as the share of generation times the unit variable cost of each technology $i$, and the fixed cost is the same multiplied by the reciprocal of the capacity factor to account for the capacity necessary to generate each technology's share of generation accounting for the technology's specific capacity factor (i.e. a low CF tech. will need more capacity to generate the same amount of electricity). The results of these equations can be found in Table B.2. The range of costs derived results in varying unit costs (or levelized cost of electricity) depending on the capacity factor, as shown in Figure B.1.

**Table B.2. Annualized fixed costs and unit terms variable costs of the synthetic baseload/backup technology used in this work**

| | $f_i$ AFC (Annualized Fixed Cost) k USD/MW per year | | | $v_b$ in unit terms vble. cost (USD/MWh) | | |
|---|---|---|---|---|---|---|
| | min | max | mean | min | max | mean |
| Gas peaking | 35 | 54 | 44 | 1.5 | 1.9 | 1.7 |
| Nuclear | 105 | 178 | 141 | 1.8 | 1.9 | 1.8 |
| Coal | 182 | 475 | 312 | 8.2 | 11.9 | 10.1 |
| Gas CC | 32 | 73 | 50 | 6.7 | 7.7 | 7.2 |
| **Synthetic** | **354** | **779** | **547** | **18.2** | **23.3** | **20.8** |

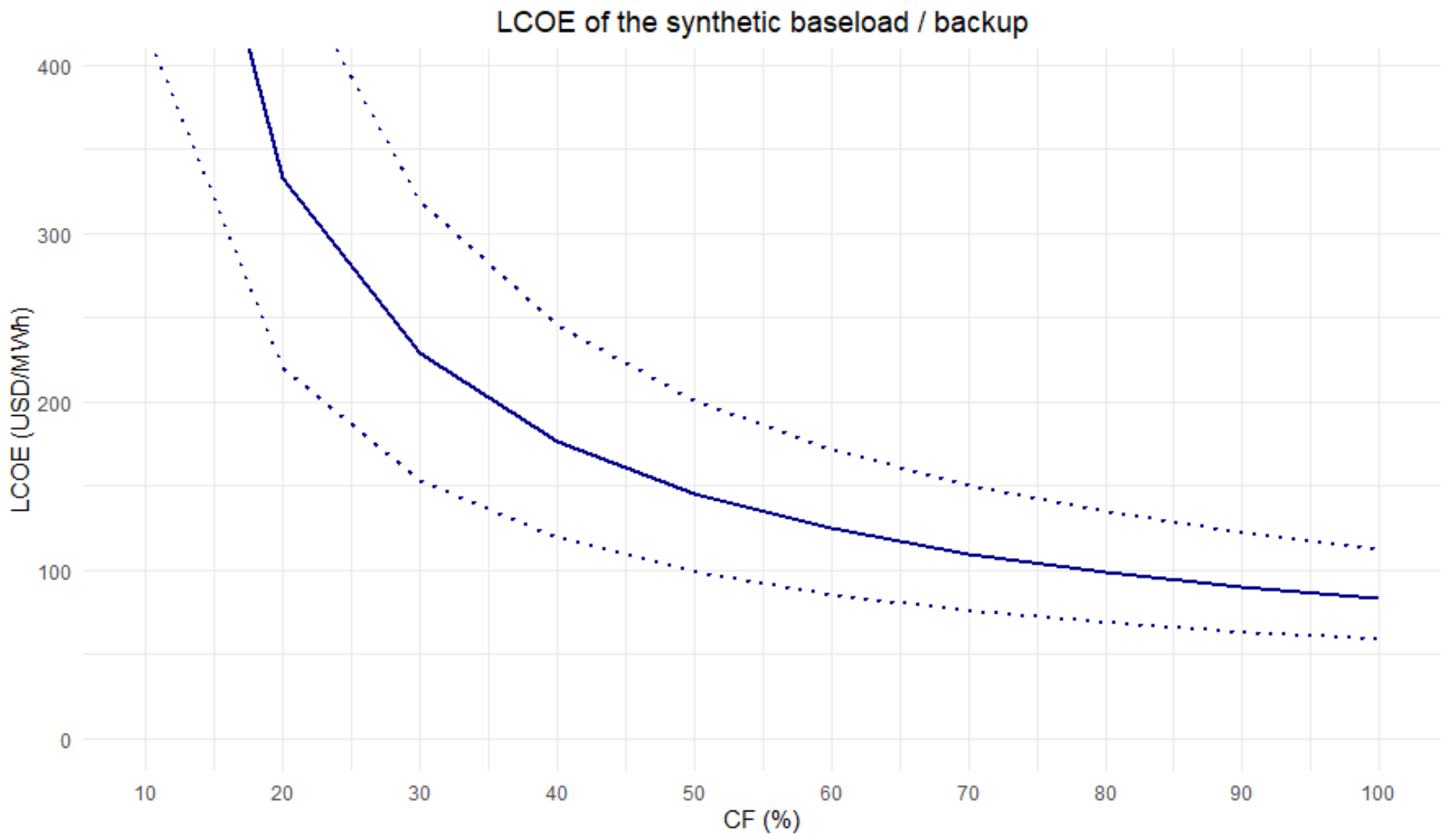

**Figure B.1. LCOE of synthetic baseload/backup mix at different capacity factors**

## B.2. Additional numerical results

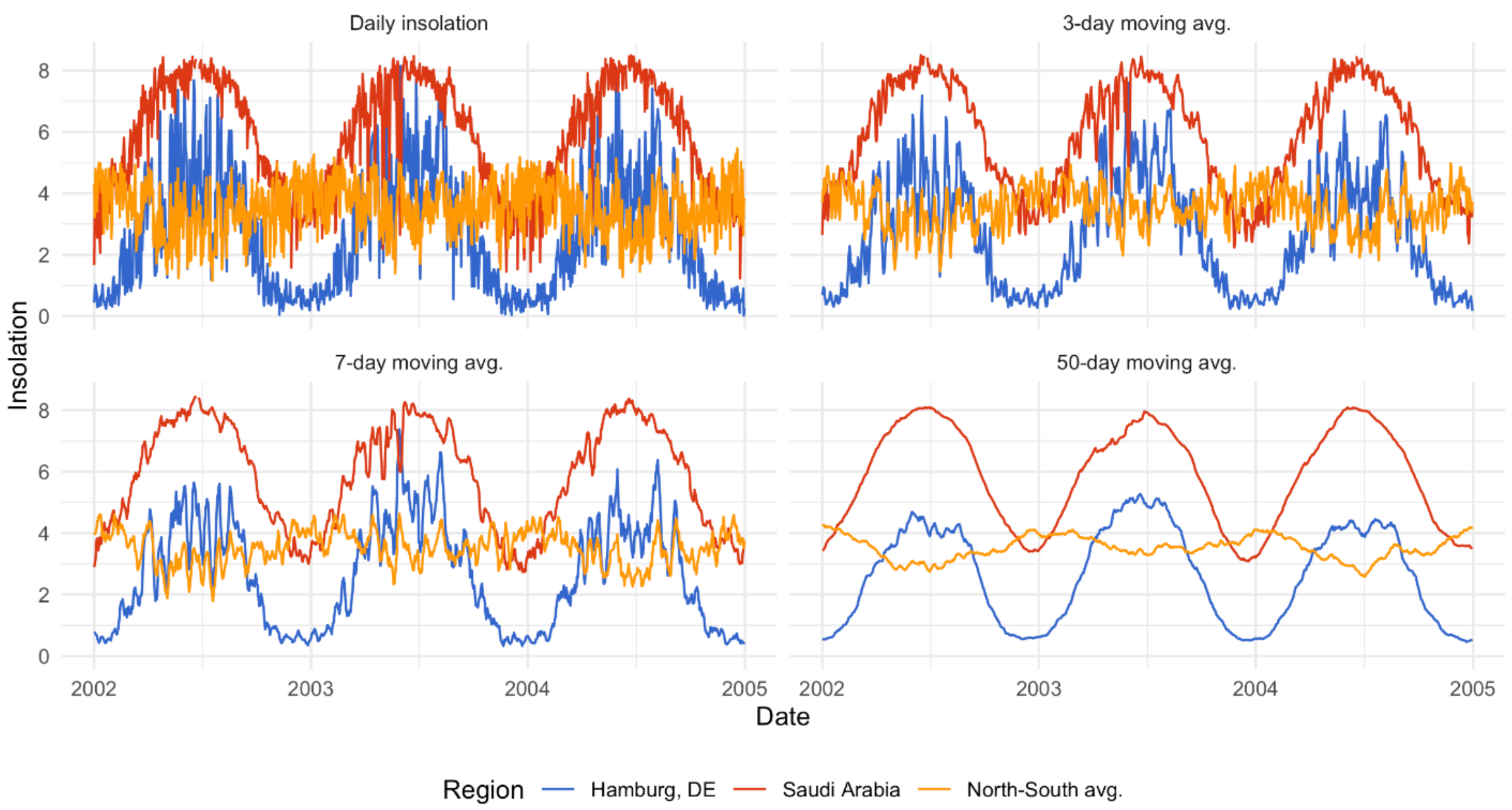

**Figure B.2. Observed insolation (in w/m²) at (i) daily, (ii) 3 day, (iii) 7-day and (iv) 50-day moving averages for Hamburg, Germany (blue line), Saudi Arabia (red) and the average for Hamburg and a location in Argentina (yellow – illustrating the elimination of the winter hole by combined North-South generation).** Using daily averages removes day-night variation, while shorter moving averages remove some intermittency due to cloud cover, but not all. A 50-day period allows for clear observation of the winter hole, the strongly reduced winter production of PV as compared to summer.

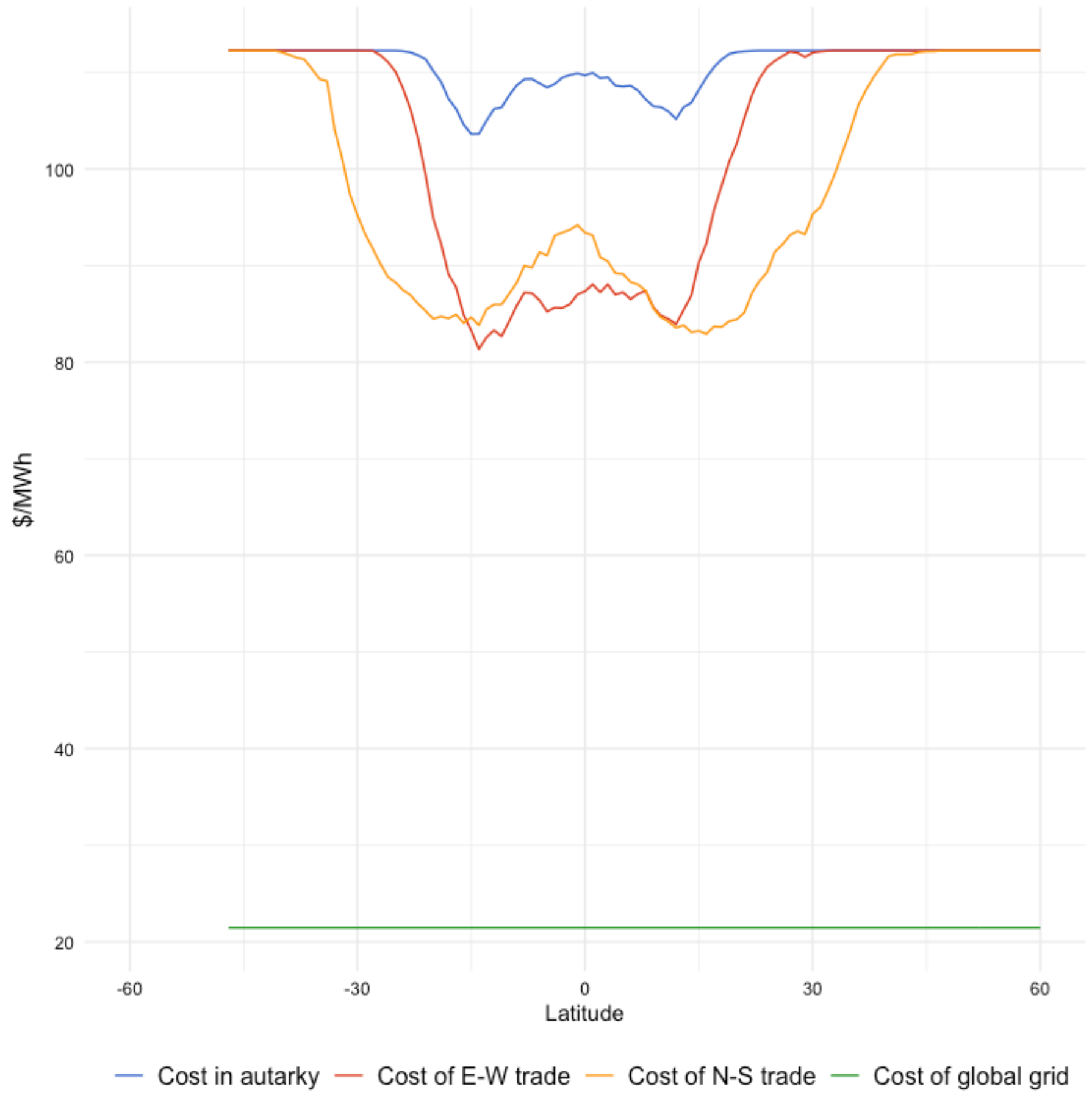

**Figure B.3. Average unit costs of electricity for autarky and all trade scenarios (assuming a high baseload/backup cost scenario) by latitude**. The gaps between the individual trade lines and the autarky line for a respective latitude illustrate the willingness to pay for transmission costs.

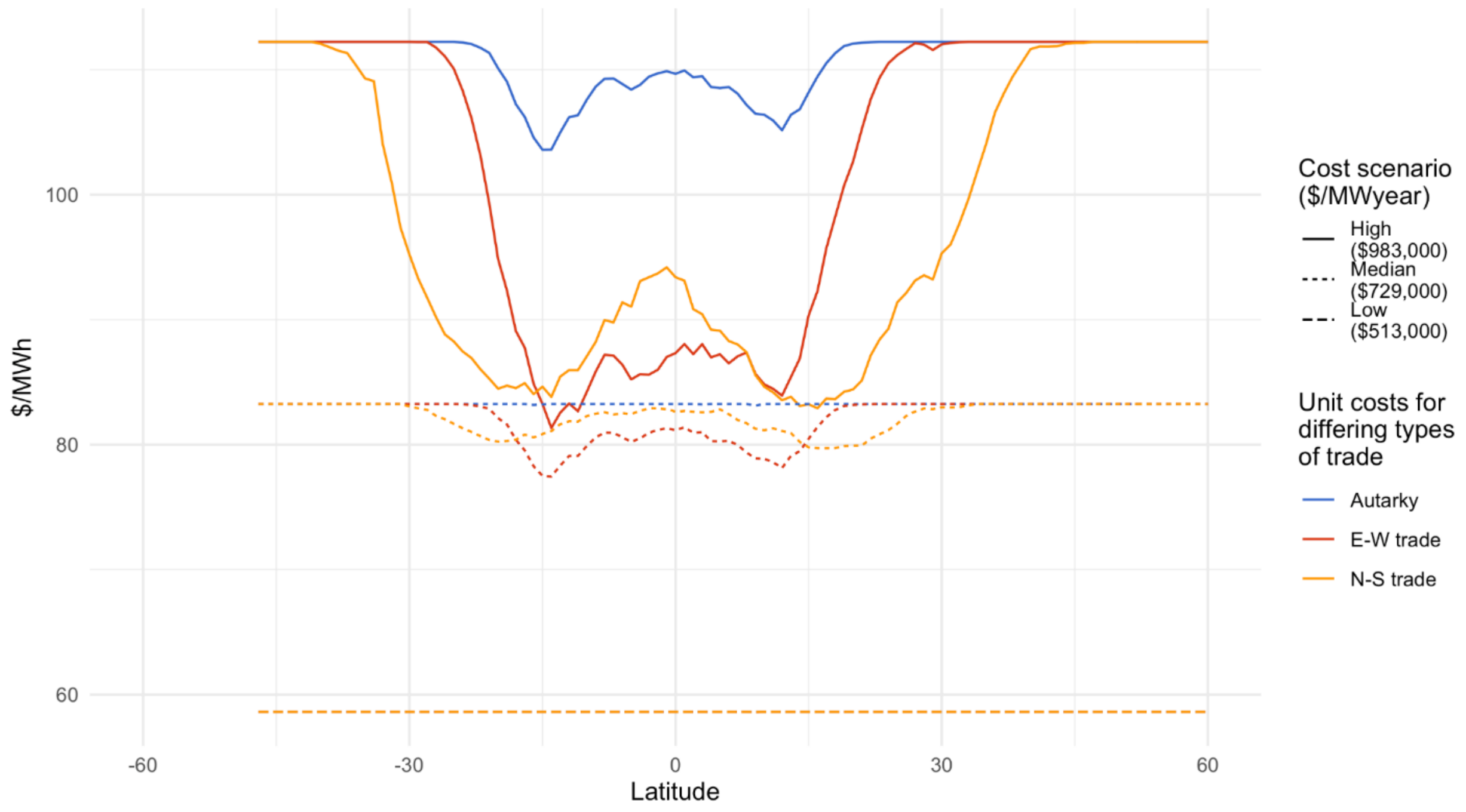

**Figure B.4. Average unit costs of electricity for autarky, East-West and North-South trade scenarios under different cost assumptions by latitude**. The gaps between the individual trade lines and the autarky line for a respective latitude illustrate the willingness to pay for transmission costs. The high cost scenario is illustrated by a solid line, median cost with a dotted line, and low with a dashed line. Note that baseload/backup costs are so low in the Low scenario as to prohibit any PV penetration in the model, thus the stacked lines just under 60 USD/MWh.

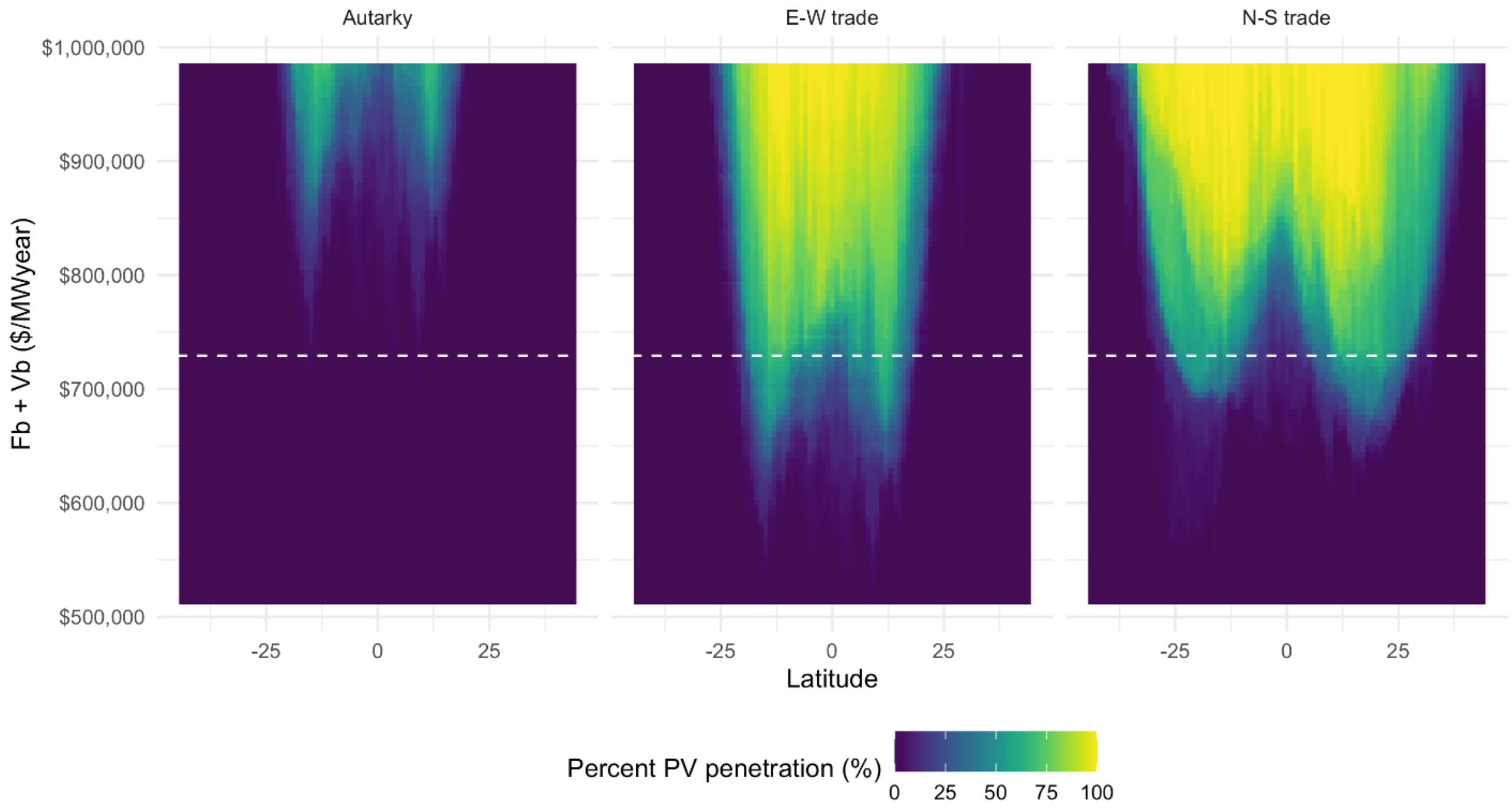

**Figure B.5. PV penetration for autarky, East-West and North-South trade regimes at varying costs of baseload/backup.** The upper and lower bounds of the plot are the costs in the high and low scenarios respectively, while the median cost scenario is indicated by the dashed line.

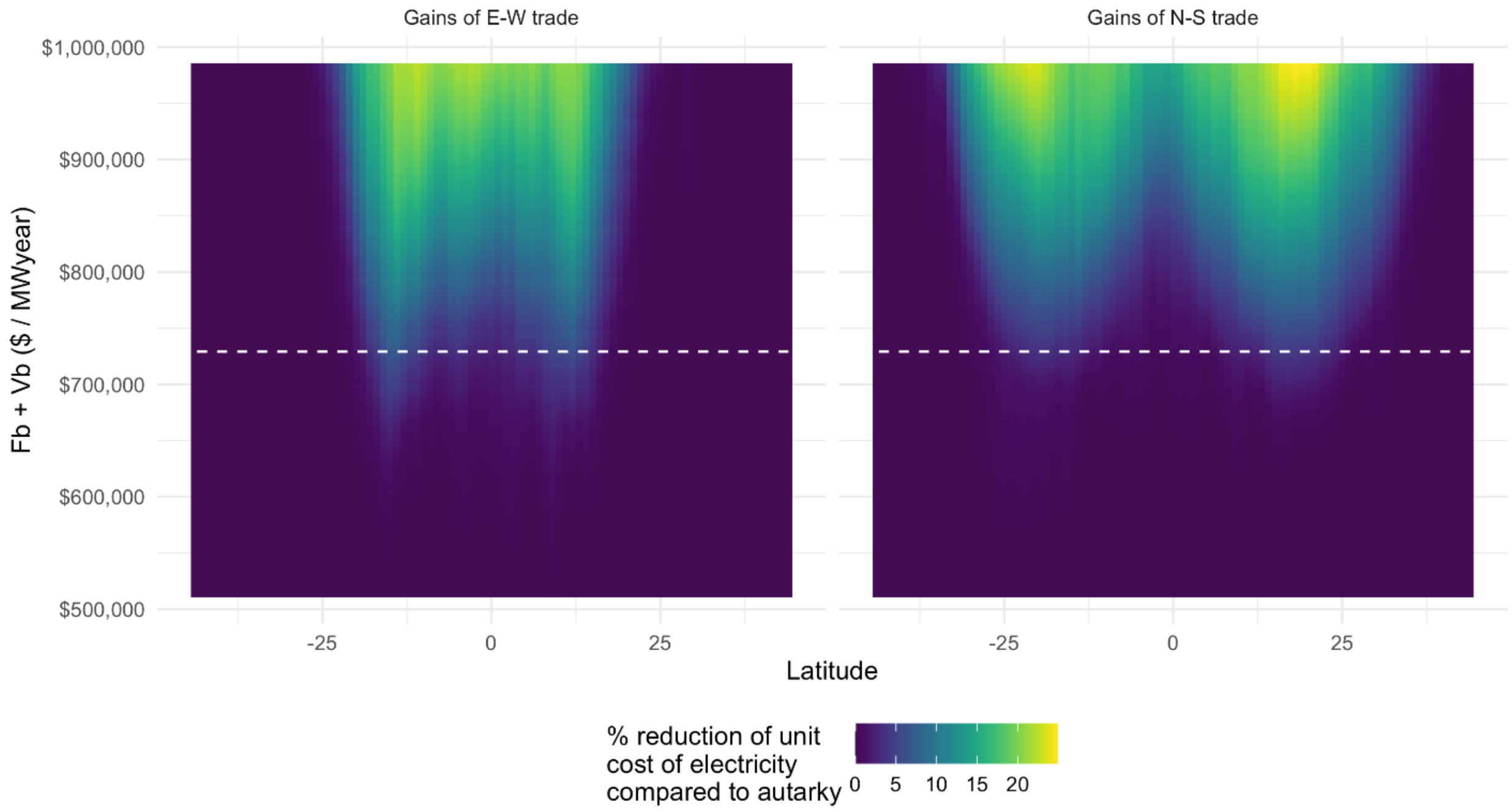

**Figure B.6. Percent reduction in unit costs of electricity in a trade case as compared to the autarky costs at varying costs of baseload/backup.** As the autarky unit costs change with varying baseload/backup costs, the willingness to pay for trade is expressed as a percentage change from autarky cost, in order to facilitate comparison across different input costs. The upper and lower bounds of the plot are the costs in the high and low scenarios respectively, while the median cost scenario is indicated by the dashed line.

## Appendix C: Transmission cost

### C.1. HVDC transmission cost in the literature

Studies have found a wide range of cost estimates for HVDC lines (see e.g. Härtel et al. (2017) and Reed et al. (2019) for a review of cost estimates and discussion). A comprehensive study of those, MacDonald et al. (2016), can be used to derive transmission costs as known in the early 2010s for the distance of the Sun Cable project, i.e. 4,500 km. For long distances, i.e. exceeding 2000 kilometers, MacDonald et al., (2016) (Supplementary Material, section 1.4 and Figure 8) indicate transmission cost at 1,000 USD/MW-mile in 2013 dollars. Assuming a transmission line life-time of 25 years, 80% utilization rate and 5% cost of capital, this translates to 28.5 USD/MWh for the distance of 4,500 kilometers[f].

**C.2. Transmission cost derived from the Sun Cable project**

Projects utilizing long-distance transmission of solar electricity that are currently under development include the Sun Cable project. This project will generate electricity from PV located in western Australia. The electricity will be transmitted through a HVDC network to Singapore. This so-called Australia-Asia link (AAPowerLink) project is seen by its initiators as a first project for the large-scale trade of electricity through extended HVDC networks from Australia "to the north", i.e. from the Australian perspective, to most of the rest of the world. The AAPowerLink will transmit green electricity over 4,500 km from Australia's Northern Territory (Newcastle Waters Station at 17.4°S, 133.40°E) to Singapore with electricity from a PV farm of 14 GW and a battery storage facility of 33 GWh, both the largest of current standards (Gordonnat and Hunt, 2020). The HVDC network consists of an undersea HVDC cable with a length of 3,770 km and an overhead line of 730 km. Announcements differ as to when this project will be operational; the official date is 2027; one of the main actors in this

[f] While simulations of PV grid configurations with storage show transmission utilization rates at about 86%, we use here a more conservative rate of 80%. A total investment cost of 6,750 M USD (4,500 km times 1.5 M USD/km for a 2.6GW line) for 25 years at 5% interest results in an annuity of 479 M USD. Dividing this by the annual electricity transmitted by the cable (16.816 M MWh) results in 28.5 USD/MWh.

undertaking, M. Cannon-Brookes, expects completion in 2025 or earlier. Electricity costs in Singapore will be 0.034 USD/kWh (Rapier, 2020).

To isolate transmission cost we first observe generation capacity. Average insolation over 20 years is 2,161 kWh/kWp in Newcastle Waters Station in a cell of 1° x 1°, i.e. an area of 111 km x 105 km at 17.5°S, 133.5°E in the Northern Territory of Australia. Average values for large areas, e.g. 1° x 1°, are usually lower than values for the best locations in such an area, so insolation could be slightly higher. At this insolation, 14 GWp of PV generate 30.25 TWh per year, i.e. an average power of 3.45 GW. The cable will transmit 2.4 GW to Singapore (Rapier, 2020). The transmission line with a full length of 4,500 km would imply an overall power loss of 14.6% (Rapier, 2020). For delivery of 2.4 GW, the cable needs 2.75 GW. The remaining 0.7 GW of electricity will be sold in Australia. Rapier (2020) calculated electricity costs of 0.034 USD/kWh by dividing the total costs of the power system by its electricity generation over 25 years. With the updated numbers given above (for example Rapier still uses a solar park of just 10 GW, not 14 GW) this would be 16.9 billion USD total costs of the project divided by (0.7+2.4) (GW of power) x 25 years x 8760 (hours/year) which is 0.025 USD/kWh. Taking into account capital costs of 5% and depreciation of 1% or 1.01 billion USD per year gives electricity costs of 0.039 USD/kWh (an update of Rapier (2020), for an upgraded system).

We calculate transmission costs for the planned AAPowerLink between Australia and Singapore. Projected total costs of the AAPower Link are 16.9 billion USD. 14 GWp of PV may cost 7 billion USD based on recent PPAs in regions with insolation (2200 kWh/kWp) like in Australia (500 USD/kWp). The battery could cost 174.7 USD/kWh at the pack level with power electronics, i.e. 5.77 billion USD for 33 GWh[g]. Subtracting both cost factors from the total costs of 16.9 billion USD gives remaining costs of 4.13 billion USD. These remaining costs pay for more than just the cable, e.g. administrative costs, costs for permits, but also additional power electronics for ancillary services. But to be on the safe side, we use the full remaining amount as cost of the transmission system. At 5% capital costs per year, 1% depreciation and 1% O&M, yearly costs of the transmission system are 289 million USD.

With a continuous transmission of 2.4 GW the transmission system delivers 21.02 TWh per year. This gives transmission costs per kWh of 289 million USD divided by 21 TWh per year or 0.0138 USD/kWh. Loss of electricity in the cable of ~3% per 1000 km, with electricity costs from the PV plant of 0.015 USD, increases transmission costs by 0.0020 USD/kWh to 0.0158 USD/kWh. At projected electricity costs of 0.039 USD/kWh in Singapore, transmission costs of 0.0158 USD/kWh make up 40.5% of total costs. Applicable more generally, for a more conservative transmission ultilization rate of only 80% this translates to 0.01975 USD/kWh.

### C.3. Further HVDC transmission cost development

Costs of submarine cables might decrease in the same pattern as costs of landlines. Costs of landlines decreased with increase of voltage (from ±600 kV to ±1100 kV), through technological progress and through economies of scale. Increase of voltage by a factor 2 increases the power by the square of 2, but would not decrease transmission costs by a factor of 4 due to higher costs of insulation for higher voltage. Assuming a cost reduction by a factor of 2 can be considered a conservative estimate. Economies of scale from production of dozens of subsea cables for very long distances might decrease manufacturing costs by another 50%. Thus transmission costs might decrease to to a quarter, i.e. 0.00396 USD/kWh for a distance of 4,500 km. Applicable more generally, for a more conservative transmission ultilization rate of only 80% this translates to 0.00495 USD/kWh. The AAPowerLink project has received early stage funding and the Australian federal government is fast-tracking it as "the world's biggest solar and storage project" (Maisch, 2020). In May 2020, Sun Cable awarded its cable route survey contract to Perth-based Guardian Geomatics.

[g] Battery costs at the pack level in 2019 were 156 USD/kWh (IEA, 2020) and were projected to decrease by 56% according to Tesla in 2020 (Musk and Baglino, 2020). As mass production of the envisaged cost-reducing battery has not yet begun we assume half of this projected cost decrease, resulting in a pack level cost of 112.3 USD/kWh. Battery packs need power electronics; for such electronics we assume 40% of present battery pack costs, i.e. 62.4 USD/kWh. Then 33 GWh of batteries cost 5.77 billion USD.

The location of this solar park and the intention to build more such parks in Australia for delivery of electricity to major load centers "in the north" is interesting from its potential to overcome intermittency of solar through connections between the two global hemispheres.

**Acknowledgements**

We thank Birgit Bednar-Friedl, Severin Borenstein, Stefan Borsky, Nebosja Nakicenovic, Thomas Schinko, Stefan Schleicher, Gernot Wagner and David Zilberman for discussion and comments on our evaluation concept.

**Funding sources**

K.W.S. and I.G. were supported by research Grant 16282 (EFFECT) of the Austrian National Bank, W.G. and K.W.S. by research Grant 776479 (COACCH), and K.W. by research Grant 837089 (SENTINEL) both under the EU Research and Innovation Framework Program Horizon 2020. J.L.P. was supported by the FWF Schrödinger fellowship Grant J 4301-G27 (Cannibalization) and by Yonsei University's Graduate School of Environmental Finance.